\documentclass[12pt, draftclsnofoot, onecolumn]{IEEEtran}

\usepackage{amsmath,epsfig,epsf,times,amssymb,booktabs,subfigure}
\usepackage{setspace}
\usepackage{array}
\usepackage{url}
\usepackage{float}
\usepackage{color}
\usepackage{graphicx}
\usepackage{amsthm,cite}
\usepackage{epstopdf}
\usepackage{enumerate}
\usepackage{multirow}
\usepackage{bbm}

\def\gap{0.65ex}
\abovedisplayskip\gap
\belowdisplayskip\gap
\abovedisplayshortskip\gap
\belowdisplayshortskip\gap

\newtheorem{lemma}{Lemma}

\theoremstyle{remark}
\newtheorem{remark}{Remark}

\newcommand{\lb}{\left(}
\newcommand{\rb}{\right)}
\newcommand{\lc}{\left\{}
\newcommand{\rc}{\right\}}
\newcommand{\cmplx}{\mathbb C}
\newcommand{\ls}{\left[}
\newcommand{\rs}{\right]}

\newcommand{\sqrtmodul}[1]{\vert\vert #1 \vert\vert}

\newcommand{\prob}[1]{{\mathbb P} \left\{ #1 \right\}}

\newcommand{\abs}[1]{ \left| #1 \right|}

\begin{document}
	
	\title{Robust Adaptive Beam Tracking for Mobile Millimetre Wave Communications}
	\author{Chunshan Liu, Min Li, Lou Zhao, Philip Whiting, Stephen~V.~Hanly, Iain~B.~Collings, Minjian~Zhao
		\thanks{Chunshan Liu and Lou Zhao are with the School of Communication Engineering, Hangzhou Dianzi University, Hangzhou, 310018, China (email: \{chunshan.liu, lou.zhao\}@hdu.edu.cn). Min Li and Minjian Zhao are with College of Information Science and Electronic Engineering, Zhejiang University, Hangzhou, 310027, China (e-mail: \{min.li, mjzhao\}@zju.edu.cn).	Philip Whiting, Stephen V. Hanly and Iain B. Collings are with the School of Engineering, Macquarie University, Sydney, NSW 2109, Australia (e-mail: \{philip.whiting, stephen.hanly, iain.collings\}@mq.edu.au).}}
	
	\maketitle
\vspace{-4em}	
\begin{abstract}
Millimetre wave (mmWave) beam tracking is a challenging task because tracking algorithms are required to provide consistent high accuracy with low probability of loss of track and minimal overhead. To meet these requirements, we propose in this paper a new analog beam tracking framework namely Adaptive Tracking with Stochastic Control (ATSC). Under this framework, beam direction updates are made using a novel mechanism based on measurements taken from only two beam directions perturbed from the current data beam. To achieve high tracking accuracy and reliability, we provide a systematic approach to jointly optimise the algorithm parameters. The complete framework includes a method for adapting the tracking rate together with a criterion for realignment (perceived loss of track). ATSC adapts the amount of tracking overhead that matches well to the mobility level, without incurring frequent loss of track, as verified by an extensive set of experiments under both representative statistical channel models as well as realistic urban scenarios simulated by ray-tracing software. In particular,  numerical results show that ATSC can track dominant channel directions with high accuracy for vehicles moving at 72 km/hour in complicated urban scenarios, with an overhead of less than 1\%.

\end{abstract}
\vspace{-1em}	
	\begin{IEEEkeywords}
	\vspace{-1em}	Beam tracking, beam refinement, high mobility, millimeter wave communications.
	\end{IEEEkeywords}
\vspace{-1.2em}	

\section{Introduction}\label{Sec:Intro}
\vspace{-0.5em}
Millimetre wave (mmWave) frequencies have been recognized as one of the most promising means to increase wireless capacity for evolving 5G and beyond cellular networks and have attracted considerable attention~\cite{pi2011introduction,andrews2014will,xiao2017millimeter,lee2018spectrum,zhang2017hybrid,liu2018decentralized,huang2018constant,8959381,8651496,yang2018low,xue2018beam}. However, employment of mmWave frequencies in mobile communications comes with significant challenges. In particular, the severe path loss in the mmWave band needs to be compensated via beamforming techniques at Base Station~(BS) and/or User Equipment (UE). To ensure high beamforming gain, the directions of the beams used in transmission must align well with the dominant path of mmWave channels at all times.

Establishing and maintaining such accurate alignment is usually obtained via an initial beam search followed by beam tracking/management~\cite{8458146}. Initial beam search on the one hand is primarily concerned with finding good beamforming directions with minimal access delay. This is generally achieved by conducting a full search over the entire angular space and hence long training times are required~\cite{liu2017Jsac,alkhateeb2015compressed}. A wide range of solutions have been proposed using techniques such as spatial scanning~\cite{xiao2017codebook,zhang2017codebook,liu2017Jsac,min2019TWC} and compressive sensing~\cite{alkhateeb2015compressed,alkhateeb2014channel,8356247}.

Beam tracking on the other hand is concerned with \emph{maintaining} high alignment accuracy between the BS/UE beams and the directions of the dominant channel paths. This is a challenging task because the directions of dominant channel paths often change rapidly due to mobility and radio environmental variations. Moreover, good mmWave beam tracking algorithms are required to (i) provide consistent high tracking accuracy and therefore high beamforming gain in various fading scenarios and (ii) use a minimum amount of tracking overhead to achieve~(i). To attain~(i), tracking errors must be kept small throughout the process of communications, not just on the average. This is because short-term large excursions of error can lead to disruptive losses of beamforming gain, with consequent packet losses or even loss of the entire link. As far as~(ii) is concerned, it is crucial that the rate of taking tracking measurements is matched to the mobility level. Using a high tracking rate on the one hand limits the time available for data transmission and tracking multiple users simultaneously, not to mention the need for UEs to track multiple BSs. Using a low tracking rate, on the other hand, will lead to frequent loss of track for high-mobility UEs, incurring costly re-alignments.

There has already been considerable effort devoted to the problem of mmWave beam tracking~\cite{80211ad,li2017analog,palacios2017tracking,yan2017wideband,zhang2016tracking,7905941,jayaprakasam2017robust,zhang2018beam,zhu2018high}. In~\cite{80211ad,li2017analog}, the search scheme of the IEEE 802.11ad standard is described. This scheme tracks mmWave channels by scanning using three beams in each tracking update: the data beam currently adopted and its two adjacent beams from a pre-determined codebook. The beam yielding the strongest measurement is selected as the next data beam which finishes the tracking udpate. This method avoids a full search of the codebook and thus reduces the tracking overhead as compared to initial alignment methods~\cite{xiao2017codebook,zhang2017codebook,liu2017Jsac,min2019TWC}. However, its accuracy is limited by the resolution of the codebooks~\cite{zhu2018high}.

In~\cite{palacios2017tracking}, it is proposed to use pilots sent in data slots to track angular changes in a hybrid mmWave beamforming architecture. Here, assumptions on the prior distributions of the angular changes are needed in order for the algorithm to operate. Reference~\cite{yan2017wideband} proposed a tracking algorithm that requires only one measurement per tracking update. This is taken at a direction perturbed with respect to the current beam direction. However, this method requires knowledge of the instantaneous strength of the path being tracked, which cannot be obtained accurately because of the unknown path angle and channel fading. References\cite{zhang2016tracking,7905941,jayaprakasam2017robust} developed Kalman Filter (KF) based algorithms to track the time-varying directions. As has been shown in~\cite{7905941,jayaprakasam2017robust}, such KF-based methods can suffer from error propagation and eventual loss of track, i.e., the angle of the dominant path is no longer covered by the beam used for data transmission.

Reference~\cite{zhang2018beam} proposed a method that requires two measurements taken at different directions in each tracking update. This method is not suitable for urban scenarios as it was developed for a mmWave UAV BS communicating with a moving ship on the sea surface, which is a line-of-sight (LOS) scenario. Furthermore, it relies on knowledge of the height of the UAV BS and the velocity of the ship to determine the measurement directions. The algorithm in~\cite{zhu2018high} takes two measurements using beams perturbed in angle with respect to the current beamforming direction. Beam updates are calculated based on the amplitude-comparison monopulse~\cite[Chapt.~9]{mahafza2017introduction}. While its main contribution lies in the consideration of hardware impairments, important algorithmic parameters such as the perturbing distance were not optimised. As we will show later, the algorithm in~\cite{zhu2018high} can suffer frequent and significant losses of beamforming gain. Moreover, none of these works provides a systematic treatment as to how much tracking overhead is needed for a given mobility level nor do they consider how to adapt the tracking overhead to different mobility levels in real-time.

In this paper, we propose a new mmWave beam tracking method which we call Adaptive Tracking with Stochastic Control (ATSC) where both BS and UE employ analog beamforming. As we will show later, ATSC is able to adapt its tracking overhead to different mobility levels, that is to use a suitable amount of overhead for a certain mobility level. Meanwhile, it can provide high tracking accuracy, as measured not only by long-term Signal-to-Noise Ratio (SNR) average but also by metrics that account for worst-case variations.

The core part of ATSC is a novel stochastic control (SC) based algorithm we developed to control the beamforming direction to achieve high alignment accuracy. In SC, two measurements taken with two sampling beams that are perturbed with respect to the direction of the current beam (data beam) are needed to get beam updates, similar to~\cite{zhu2018high,zhang2018beam}. By analysing the statistical behaviour of the measurement signals, we design a simple yet effective beam updating mechanism. This mechanism enables us to analytically derive upper bounds on the probability of losing tracking (PLT) as the channel changes, i.e., the probability that the pointing error goes outside a given fraction of the beam width between two updates. In order to achieve consistently high tracking accuracy in both high-mobility and low-mobility scenarios, we further optimise the parameters for the SC algorithm,  including the perturbing distance, pilot sequence length and the stepsize of SC update, based on the derived PLT upper bound and the mean absolute error (MAE) between the angle of the dominant path and the beam angle updated by SC.

Finally, we introduce a method to adaptively set the tracking rate according to the underlying speed of angular variations. This angular speed can be estimated directly from the angle updates produced by SC, owing to its high tracking accuracy. With the SC algorithm and the adaptive tracking rate,  ATSC is able to maintain high beam alignment accuracy without suffering error propagation. It adaptively reduces the tracking rate (hence the overhead) when the dominant path angle varies slowly to save overhead, and increases the tracking rate when the angle varies quickly (due to faster mobility). The performance of the proposed ATSC protocol is evaluated via an extensive set of numerical results, including performance evaluations in complicated urban scenarios simulated via high-accuracy ray-tracing software~\cite{Insite}.

The remainder of the paper is organized as follows. Section \ref{Sec:System} presents the system model and describes the beam tracking protocol at high-level. Section \ref{Sec:Control} elaborates the SC algorithm, the core of ASTC, while Section~\ref{Sec:Algorithm} details the full ASTC framework. Numerical simulations are presented in Section~\ref{Sec:Results} and conclusions are finally drawn in Section~\ref{Sec:Conclusions}.
	
\vspace{-0.5em}	
\section{System Model and Beam Tracking Protocol \label{Sec:System}}
\vspace{-0.5em}	
\subsection{Beam Tracking Protocol}\label{Sec:protocol}
\begin{figure}
	\centering
	\vspace{-1.5em}
	\includegraphics[width=0.95\textwidth]{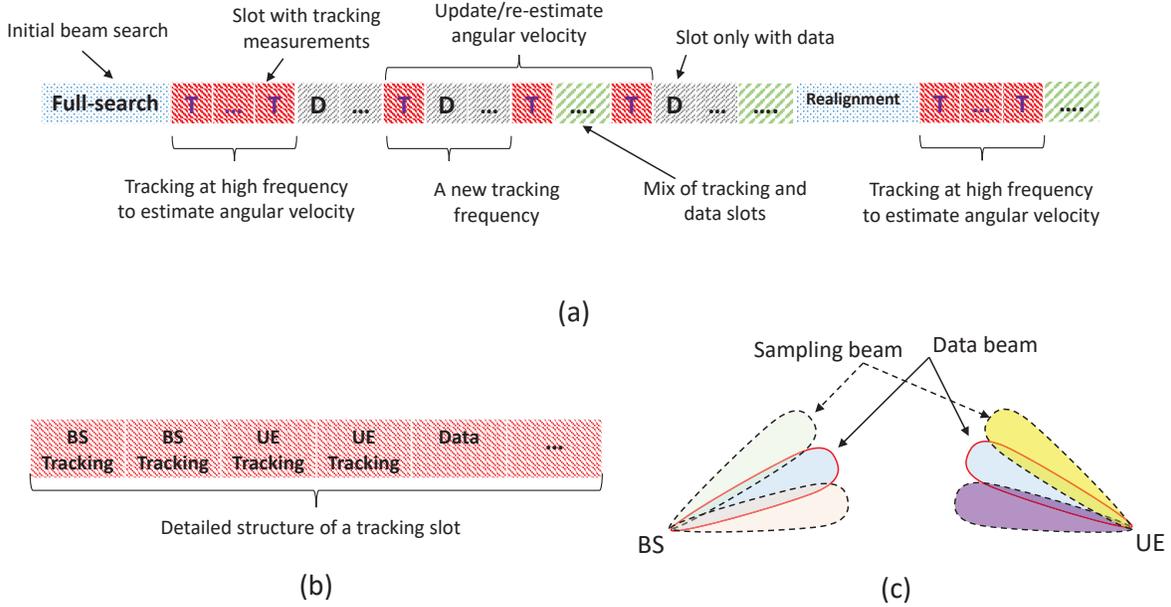}
	\vspace{-1.2em}
	\caption{An illustration of the proposed mmWave beam tracking strategy: (a) A possible frame structure; (b) Detailed structure of a tracking slot; (c) An illustration of beam tracking measurement.\label{fig_system}}
	\vspace{-1.5em}
\end{figure}

\vspace{-0.5em}	
We consider a mmWave communication system where a BS and a UE are both equipped with  one-dimensional Uniform Linear Arrays (ULAs) with isotropic antennas, a single RF chain and both use analog beamforming. The BS and UE attempt to track the beams used for data transmission such that the data beams are always aligned well with the time-varying dominant path of the channel.
Fig.~\ref{fig_system} (a) is a time frame structure of our proposed tracking protocol. The time frame consists of times slots with beam tracking measurements, i.e., tracking slots, and time slots without tracking measurements, i.e., data slots. As illustrated in Fig.~\ref{fig_system} (a), the tracking operations start after a full search over the entire angular space, where the full-search corresponds to an initial beam alignment~\cite{min2019TWC} when a UE wakes up from idle mode. Immediately after the full-search, tracking measurements are collected frequently, i.e., every time slot becomes a tracking slot. In each of these slots, the beam used for data transmission, i.e., the \emph{data beam}, is updated. After a number of tracking updates, the angular speed is estimated, based on which, the tracking rate (how frequently tracking updates are performed) is determined so as to match the tracking overhead to the mobility level. To deal with the possibilities of sudden blockage of the existing path being tracked and failure to track the path of interest, realignment is also considered in the proposed protocol, see Fig.~\ref{fig_system}(a).

Fig.~\ref{fig_system}(b) gives a more detailed view of a tracking slot, where the tracking measurements take a fraction of the slot, leaving the remaining parts for other uses such as data transmission. To track the beam at either the BS or the UE, two measurements are collected, as indicated in Fig.~\ref{fig_system} (b). If both the beams at the BS and the UE are to be tracked, four measurements are needed in total for ULA beams: two consecutive measurements at the BS and two consecutive measurements at the UE. For each side, the two measurements are taken by measuring pilot sequences, sent by the other side using the current data beam. The two measurements are collected using two \emph{sampling beams} which point in two directions that are slightly perturbed with respect to the direction of the data beam used prior to the tracking measurements, as illustrated in Fig.~\ref{fig_system}~(c).

\begin{remark}
The above mentioned protocol can be generalised to track multiple users/beams. Suppose the users are served in a time-division manner. Then the tracking measurements of different users can be scheduled to be collected at different times. For instance, a tracking slot now can contain tracking measurements from different users such that the BS can update its beam directions for the users. Similar treatments can be made to track multiple paths.  As we will show later, the overhead due to tracking a single path for one user is small. This makes it possible to support more complicated scenarios including multi-user, multi-BS and multi-path.~$\square$
\end{remark}

Before proceeding to Section III to elaborate the control algorithm, we present the channel model and the signal model in the remaining part of this section.

\vspace{-0.8em}	
\subsection{Channel Model}
\vspace{-0.5em}	
We consider a block fading model where the channel remains fixed during a slot. Denote $H(t)\in \cmplx^{N_R \times N_T}$ as the channel matrix between the BS and the UE at slot $t$, where $N_T$ and $N_R$ are the number of antennas at the BS and the UE, respectively. Following a geometric model, $H(t)$ can be represented as:
\begin{equation}\label{eq:channel_model}
H(t) = \sum_{l=1}^L g_l(t)\mathbf{u}(\tilde{\phi}_l(t))\mathbf{v}^\dag(\tilde{\psi}_l(t)),
\end{equation}
where $L$ is the number of multipath components, $g_l(t)$ is the complex coefficient of the $l$-th path in slot $t$, and $\tilde{\phi}_l(t)$ and $\tilde{\psi}_l(t)$ are the angles of the $l$-th path with respect to the UE and the BS, respectively. Vectors $\mathbf{u}(\phi)$ and $\mathbf{v}(\psi)$ are the array response vectors at the UE and the BS, respectively. For ULA, it is often convenient to consider $\phi\doteq sin(\tilde{\phi})$ and $\psi\doteq sin(\tilde{\psi})$. Using these notations, the array response vectors can be represented as:
$$
\mathbf{u}(\phi) = \ls  1, e^{ -j 2\pi \frac{d}{\lambda} \phi}, \cdots,   e^{-j 2\pi \frac{d}{\lambda} (N_R - 1) \phi} \rs^T
~\text{and}~
\mathbf{v}(\psi) = \ls  1, e^{-j 2\pi \frac{d}{\lambda} \psi}, \cdots,   e^{-j 2\pi \frac{d}{\lambda} (N_T- 1) \psi} \rs^T,
$$
where $(\cdot)^T$ denotes vector transpose, $d$ is the antenna spacing and $\lambda$ the carrier wavelength.

\vspace{-1em}
\subsection{Signal Model and Tracking Measurements}
\vspace{-0.5em}
We go on to present the signal model and explain how measurements are obtained under our tracking algorithm. Denote the pilot sequence used in the beam tracking process as $\mathbf{s}\in \cmplx^{n\times 1}$, where $\|\mathbf{s}\|_2^2=nP_T$ with $P_T$ being the transmit power. As explained in Section~\ref{Sec:protocol}, for the BS to track the beam, the UE sends the pilot sequence twice in the uplink using the UE data beam used in the previous time slot, while the BS takes the corresponding measurements using two beamformers that are perturbed slightly with respect to the BS data beam. The same process applies in the downlink such that the UE updates its data beam. We consider ULA beamformers at both the UE and the BS for both the data beams and the sampling beams. UE beam $\mathbf{w}(\Phi)$ pointing in direction $\Phi$ and BS beam $\mathbf{f}(\Psi)$ pointing in direction $\Psi$ can be represented as:{\small
$$
\mathbf{w}(\Phi) = \frac{1}{\sqrt{N_R}}\ls  1, e^{ -j 2\pi \frac{d}{\lambda} \Phi}, \cdots,   e^{-j 2\pi \frac{d}{\lambda} (N_R - 1) \Phi} \rs^T
~\text{and}~
\mathbf{f}(\Psi) = \frac{1}{\sqrt{N_T}}\ls  1, e^{ -j 2\pi \frac{d}{\lambda} \Psi}, \cdots,   e^{-j 2\pi \frac{d}{\lambda} (N_T- 1) \Psi} \rs^T.
$$}
We note that the ULA beams achieve the highest possible gain if the beamformer direction is aligned perfectly with the path direction, assuming there is only one path.

Let $\Psi(t)$ and $\Phi(t)$ be the directions of the data beams for the BS and UE, respectively, and $\Delta_T$ and $\Delta_R$ be the angular perturbation applied at the BS and the UE, respectively. In the $t$-th slot, for BS beam tracking, two measurements are taken with beamformers pointing to $\psi^{\pm}(t) \doteq \Psi(t-1)\pm \Delta_T$. Similarly, we can define $\phi^{\pm}(t) \doteq \Phi(t-1)\pm \Delta_R$ for UE beam tracking. Denote further
\begin{equation}
\text{Track BS beam:}~~h_T^{\pm} = \mathbf{w}^\dag(\Phi(t-1))\mathbf{H}(t)\mathbf{f}(\psi^{\pm}(t))
\label{eqn_psi}
\end{equation}
and
\begin{equation}
\text{Track UE beam:}~~h_R^{\pm} = \mathbf{w}^\dag(\phi^{\pm}(t))\mathbf{H}(t)\mathbf{f}(\Psi(t-1))
\label{eqn_phi}
\end{equation}
as the effective channels of the BS and the UE measurements, respectively. The measurements collected by the BS and the UE can be represented as:
\begin{equation}\label{eqn_ypsi}
\mathbf{y}_T^{\pm}(t) = h_T^{\pm}(t)\mathbf{s} + \mathbf{z}^{\pm}_T(t),~~~~\text{and}~~~~\mathbf{y}_R^{\pm}(t) = h_R^{\pm}(t)\mathbf{s} + \mathbf{z}^{\pm}_R(t),
\end{equation}
respectively, where $\mathbf{z}^{\pm}_T$ and $\mathbf{z}^{\pm}_R$ are vectors of circularly symmetric Gaussian random variables with zero mean and variance $\sigma^2$. By matched filtering the measurements with pilot squence $\mathbf{s}$, we obtain
\begin{equation}\label{eq_Qr_Qt}
Q_T^{\pm}(t) = \frac{2}{\sigma^2\sqrtmodul{{\bf s}}^2} \left |{\bf s}^\dagger \mathbf{y}_T^{\pm}(t)\right |^2,~~~~\text{and}~~~~
Q_R^{\pm}(t) = \frac{2}{\sigma^2\sqrtmodul{{\bf s}}^2} \left|{\bf s}^\dagger \mathbf{y}_R^{\pm}(t)\right|^2.
\end{equation}
It follows directly from \eqref{eqn_ypsi} that $Q_T^{\pm}(t)$ and $Q_R^{\pm}(t)$ are independent non-central chi-square ($\chi^2$) random variables with Degrees of Freedom (DOF)  equal to 2 and non-centrality parameters
\begin{equation}\label{eqn_eta}
\eta_T^\pm(t)  = \frac{2 nP_T\abs{h_T^\pm(t)}^2}{\sigma^2},~~~~\text{and}~~~~\eta_R^\pm(t) = \frac{2 nP_T\abs{h_R^\pm(t)}^2}{\sigma^2}.
\end{equation}
The above results are used in the derivation of the update formula for our SC algorithm.

\vspace{-1em}
\section{Stochastic-Control Algorithm for Beam Tracking\label{Sec:Control}}
\vspace{-0.5em}
In this section, we present the core part of the proposed beam tracking framework, the stochastic-control (SC) algorithm. For ease of exposition, we use beam tracking at the BS side to describe the algorithm, assuming that the UE beam is fixed and does not require tracking. We also assume that the channel has a single path when developing the SC algorithm, i.e., $L=1$, and drop the subscript $l$ for convenience. We note that the single-path channel model assumption is reasonable  because narrow beams are used for tracking and mmWave channels tend to be sparse in the angular domain~\cite{Akdeniz2014}. In this case, the chance that two or more channel paths falling into the same BS and UE beam is small~\cite{zhu2018high}. Therefore, the single-path model provides sufficient accuracy even in the presence of multiple paths. We finally assume in the section that the path coefficient~$g$ does not change over time\footnote{In the numerical experiments, we will examine the performance of the proposed algorithm using multi-path fading channels.} and the error between the UE data beam and the angle at the UE is constant: $\phi(t) - \Phi(t) = \epsilon_R$.

Under the assumptions made above, the effective channel gains of the two BS measurements can be represented as:
\begin{align}
|h_T^{\pm}(t)|^2 & = |\mathbf{w}^\dag(\Phi(t-1))\mathbf{H}(t)\mathbf{f}(\psi^{\pm}(t))|^2 = |g\mathbf{w}^\dag(\Phi(t-1))\mathbf{u}(\phi(t))\mathbf{v}^\dag(\psi(t))\mathbf{f}(\psi^{\pm}(t))|^2 \\
& = |g|^2G_R(\epsilon_R)G_T(\Psi^{\pm}(t)-\psi(t))
\end{align}
where $G_T(\Psi^{\pm}(t) - \psi(t))=|\mathbf{v}^\dag(\psi(t))\mathbf{f}(\Psi^{\pm}(t))|^2= \left|\frac{1}{\sqrt{N_T}}\sum_{k=0}^{N_T-1}e^{j2\pi\frac{d}{\lambda}(\Psi^{\pm}(t)-\psi(t))}\right|^2$ are the gains of the two sampling beams at the channel direction. $G_R(\epsilon_R)$ is defined similarly. Clearly, $G_T(0) = N_T$. Moreover, $G_T(\epsilon)/G_T(0) = G_T(\epsilon)/N_T$ only depends the relative error $\epsilon/B_T$.

\begin{figure}
	\centering
	\vspace{-1.5em}
	\includegraphics[width=0.5\textwidth]{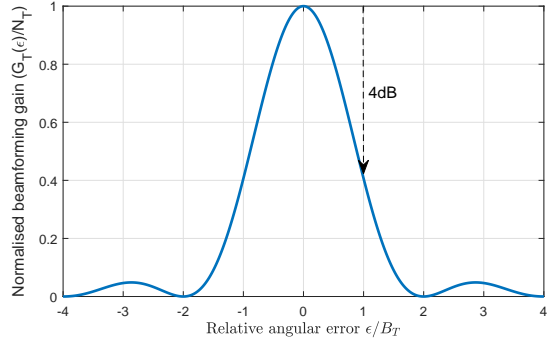}
	\caption{Normalised beamforming gain $G_T(\epsilon)/N_T$ with respect to relative beam error $\epsilon/B_T$: ULA with isotropic antennas. \label{fig_ULA64}}
	\vspace{-1.5em}
\end{figure}

Fig.~\ref{fig_ULA64} plots $G_T(\epsilon)/N_T$ with respect to the relative error $\epsilon/B_T$. It can be seen that $G_T(\pm 2B_T)\approx 0$ and $G_T(\pm B_T)$ is about $4$ dB lower than $G_T(0)$. A larger relative angular error will lead to even more significant loss of beamforming gain, which is to be avoided in beam tracking.

Let $\epsilon_\psi(t) = \Psi(t-1)-\psi(t)$ be the initial angular error prior to the tracking measurements at slot $t$. Since $\Psi^{\pm}(t) = \Psi(t-1) \pm \Delta_T$, $\Psi^{\pm}(t)-\psi(t) = \epsilon_\psi(t)\pm \Delta_T$.  With the notations defined above, the non-centrality parameters of $Q^{\pm}_T(t)$ in~\eqref{eqn_eta} can be further represented as:
\begin{equation}\label{eq:non_centrality}
\eta^\pm(t) = \frac{2 nP_T|g|^2G_R(\epsilon)G_T(\epsilon_\psi(t)\pm \Delta_T)}{\sigma^2} = 2n\gamma G_R(\epsilon_R)G_T(\epsilon_\psi(t)\pm \Delta_T),
\end{equation}
where $\gamma = P_T|g|^2/\sigma^2$ is the SNR of the dominant path without beamforming.

\vspace{-1em}
\subsection{Stochastic-Control algorithm outline}
\vspace{-0.5em}
Consider the outcome of an exhaustive initial search based on a set of equally spaced ULA beams~\cite{liu2017Jsac,min2019TWC}, with the space between two beam centres equal to $2/N_T = 2B_T$ in the $sin(\tilde{\psi})$ domain. A successful initial search finds the best beam from the codebook, whose angular difference from the channel is smaller than that of any other beams in the codebook. In this case, the angular error (after a successful initial search) is within $[-B_T,+B_T]$ and the beamforming gain will be close to the maximal gain that can be achieved.

In the process of beam tracking, it is also desirable to maintain the angular error within $[-B_T,+B_T]$. Suppose this goal is achievable, then it is reasonable to consider that
prior to a tracking update, $\epsilon_\psi(t) \in [-B_T,+B_T]$. The control algorithm  will then update the beam angle from $\Psi(t-1)$ to $\Psi(t)$ at slot $t$, ideally
 to reduce the post-tracking error
\begin{equation}\label{eq:error_update1}
 \epsilon'_\psi(t) = \Psi(t)-\psi(t)
 \end{equation}
 such that $|\epsilon'_\psi(t)|<|\epsilon_\psi(t)|$.

The two measurements $Q^{\pm}_T(t)$ provide useful information to fulfil this goal. To see this, consider $\epsilon_\psi(t)>0$ without loss of generality, then $|\epsilon_\psi(t)+\Delta_T|>|\epsilon_\psi(t)-\Delta_T|$. If $\epsilon_\psi(t)\pm\Delta_T$ are not too far from $0$, i.e., being still in $[-2B_T,2B_T]$, then it can be seen that $\eta^-_T(t)>\eta_T^+(t)$.\footnote{If $|\epsilon_\psi(t)\pm\Delta_T|>2B_T$, it can be seen that $\eta^{\pm}_T(t)$ is close to zero as the beamforming gain is very small.} With a sufficiently long pilot sequence, i.e., $n$ is large enough, there is a high chance that $Q^{-}_T(t)-Q^{+}_T(t)>0$. In this case, to achieve the goal that $|\epsilon'_\psi(t)|<|\epsilon_\psi(t)|$, the angular update can take the form:
$\Psi(t) = \Psi(t-1) + h(Q_T^+(t)-Q^-_T(t))$,
where $h(x)$ is a function that $h(x)>0$ if $x>0$ and $h(x)<0$ if $x<0$. With this update, the post-tracking error becomes $\epsilon'_{\psi}(t)=\epsilon_{\psi}(t)+ {h}(Q_T^+(t)-Q^-_T(t))$ and will be reduced in magnitude if $h(\cdot)$ is properly chosen.

In this paper, we take a simple yet effective choice of $h(x)$: $h(x) = \delta x$, where $\delta>0$ is a stepsize parameter controlling the size of the update/correction. Following this choice, $Q_T^+(t)-Q^-_T(t)$ itself is a random variable,  whose expectation is the difference between the two corresponding non-centrality parameters
$$\eta_T^+(t) - \eta_T^-(t) = 2n\gamma  G_R(\epsilon_R)[G_T(\epsilon_\psi(t)+\Delta_T)-G_T(\epsilon_\psi(t)-\Delta_T)],$$
as can be seen from \eqref{eq:non_centrality}. As the channel is being tracked and data transmitted, it is reasonable to consider that an accurate estimate of the post-beamforming SNR prior to the tracking measurements at slot $t$ is available, i.e., $\Gamma(t-1) = 2n\gamma G_R(\epsilon_R)G_T(\epsilon_\psi(t))$. Note that the path strength, or equivalently the pre-beamforming SNR $\gamma$ is generally not available due to the unknown angular error $\epsilon_R$ and $\epsilon_{\psi}(t)$.

Using $\Gamma(t-1)$ to normalise  $Q_T^+(t)-Q^-_T(t)$, we obtained a modified $h(Q_T^+(t)-Q^-_T(t))$ as follows:
\begin{equation}\label{eq:linear_drift}
h(Q_T^+(t)-Q^-_T(t)) = \delta_T \frac{Q_T^+(t)-Q^-_T(t)}{\Gamma(t-1)},
\end{equation}
where we call function $h(\cdot)$  the drift function and $\delta_T>0$ is a stepsize.

Before continuing, we note that for any choice of the stepsize $\delta_T>0$, it is possible that $|h(Q_T^+(t)-Q^-_T(t))|>B_T$, because $Q_T^{\pm}$ are random. When $|h(Q_T^+(t)-Q^-_T(t))|>B_T$, an over correction occurs because the initial error $\epsilon_{\psi}(t) \in [-B_T,+B_T]$. We therefore apply a truncation to $h(Q_T^+(t)-Q^-_T(t))$ and obtain the following beam updating formula for the control algorithm
\begin{equation}\label{eq:track_update}
\Psi(t) = \Psi(t-1) + \tilde{h}(Q_T^+(t)-Q^-_T(t)),
\end{equation}
where
\begin{equation}
\tilde{h}(Q_T^+(t)-Q^-_T(t)) = \min \lb  \abs{h(Q_T^+(t)-Q^-_T(t))}, B_T \rb \cdot z(h(Q_T^+(t)-Q^-_T(t))),		
\label{eqn_control}								
\end{equation}
where $z(\cdot) \in \lc -1, 1 \rc$ is the sign function. With tracking update $\tilde{h}(Q_T^+(t)-Q^-_T(t))$, the post-tracking error defined in~\eqref{eq:error_update1} becomes:
\begin{equation}\label{eq:error_update}
\epsilon'_{\psi}(t) = \epsilon_{\psi}(t)+ \tilde{h}(Q_T^+(t)-Q^-_T(t)),
\end{equation}
Eq.~\eqref{eqn_control} specifies the control algorithm except that the sampling angle $\Delta_T$ and the step size $\delta_T$ have not yet been specified.

\begin{figure}[t]
	\vspace{-1.5em}
	\centering
	\includegraphics[width = 0.5\textwidth]{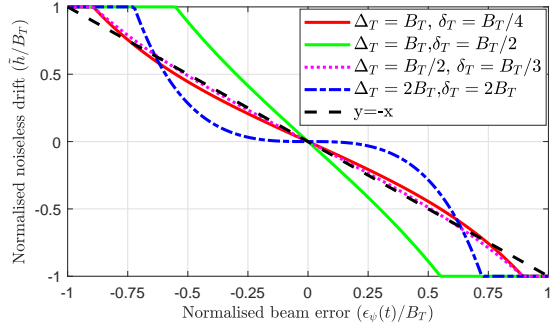}
	\vspace{-1.2em}
	\caption{Truncated drift functions for four parameter pairs when $n\rightarrow \infty$. \label{fig_DriftPlots}}
	\vspace{-1.5em}
\end{figure}

\begin{remark}
The linear drift given~\eqref{eq:linear_drift} is motivated by considering $Q_T^+(t)-Q^-_T(t)$ in the noiseless case ($n\rightarrow \infty$), where $Q_T^+(t)$ and $Q^-_T(t)$ become deterministic and so does the drift function:  $h = \delta_T \cdot\frac{G_T(\epsilon_{\psi}(t)+ \Delta_T)  - G_T(\epsilon_{\psi}(t)- \Delta_T)}{G_T(\epsilon_{\psi}(t))}$. With proper choices of parameter $\Delta_T$ and $\delta_T$, the noiseless drift after truncation, i.e., $\tilde{h}$, is close to $-\epsilon_{\psi}(t)$. In this case, following~\eqref{eq:error_update}, the error after update will be small: $\epsilon'_{\psi}(t) = \epsilon_{\psi}(t)+\tilde{h}\approx\epsilon_{\psi}(t)-\epsilon_{\psi}(t)= 0$. Fig.~\ref{fig_DriftPlots} plots the drift given by~\eqref{eqn_control} when $n\rightarrow\infty$, with respect to the initial error $\epsilon_{\psi}(t)$. Four parameter choices are presented: $(\Delta_T = B_T,\delta_T=B_T/4)$, $(\Delta_T = B_T/2,\delta_T=B_T/3)$, $(\Delta_T = B_T,\delta_T=B_T/2)$ and $(\Delta_T = 2B_T,\delta_T=2B_T)$. The drift and the initial error plotted are normalised by $B_T$ such that they are in $[-1,1]$. It can be seen that the two parameter choices $(\Delta_T = B_T,\delta_T=B_T/4)$ and $(\Delta_T = B_T/2,\delta_T=B_T/3)$ provide a drift close to the curve $y=-x$. However, with $(\Delta_T = B_T,\delta_T=B_T/2)$ or $(\Delta_T = 2B_T,\delta_T=2B_T)$, the gap between the drift and $y=-x$ is fairly large, leading to higher expected error after tracking updates. This result shows that it is possible to achieve low tracking error by taking a linear drift. It also shows that parameter $\Delta_T$ and $\delta_T$ are crucial to the accuracy of the tracking algorithm. $\square$
\end{remark}

\vspace{-1em}			
\subsection{Choosing the Control Parameters}
\vspace{-0.5em}
In this subsection, we investigate the impact of the parameter pair $(\Delta_T,\delta_T)$, together with the pilot sequence length $n$, on the performance of the tracking algorithm. We consider two important performance metrics: the mean absolute error (MAE) and the probability of losing track (PLT) between two tracking corrections. The MAE is an expectation of the alignment error immediately after the tracking updates, i.e., $\mathbb E\{|\epsilon'_{\psi}(t)|\}$, assuming that the initial error $\epsilon_{\psi}(t)$ is random. To define PLT, we will say that the algorithm has lost track if the beam pointing error is outside the interval $[-B_T,B_T]$, i.e., the channel direction is no longer covered by the mainlobe of the data beam. In what follows, we will use these metrics to choose the parameters.

\subsubsection{Mean Absolute Error $\mathbb E\{|\epsilon'_{\psi}(t)|\}$}
In computing the MAE, we consider uniform random $\epsilon_{\psi}(t)$ in $[-B_T,+B_T]$ which corresponds to the error distribution immediately after the initial alignment or a realignment where an exhaustive search based on a fixed codebook is used and the path angle is uniformly distributed in $[-1,1]$ (see~\cite{min2019TWC} as an example). The MAE captures the tracking accuracy immediately after tracking updates and also reflects the tracking accuracy in static scenarios where angles do not change after beam updates. We note that the tracking accuracy in static scenarios is an important performance metric for the beam tracking algorithm, because a good tracking algorithm should not only be able to track angles when there are changes, but also be able to keep the angle close to the true one when there are no changes.

\begin{figure}
	\vspace{-1.5em}
	\centering
	\begin{minipage}{.48\linewidth}
		\centering
		\includegraphics[width=1\textwidth]{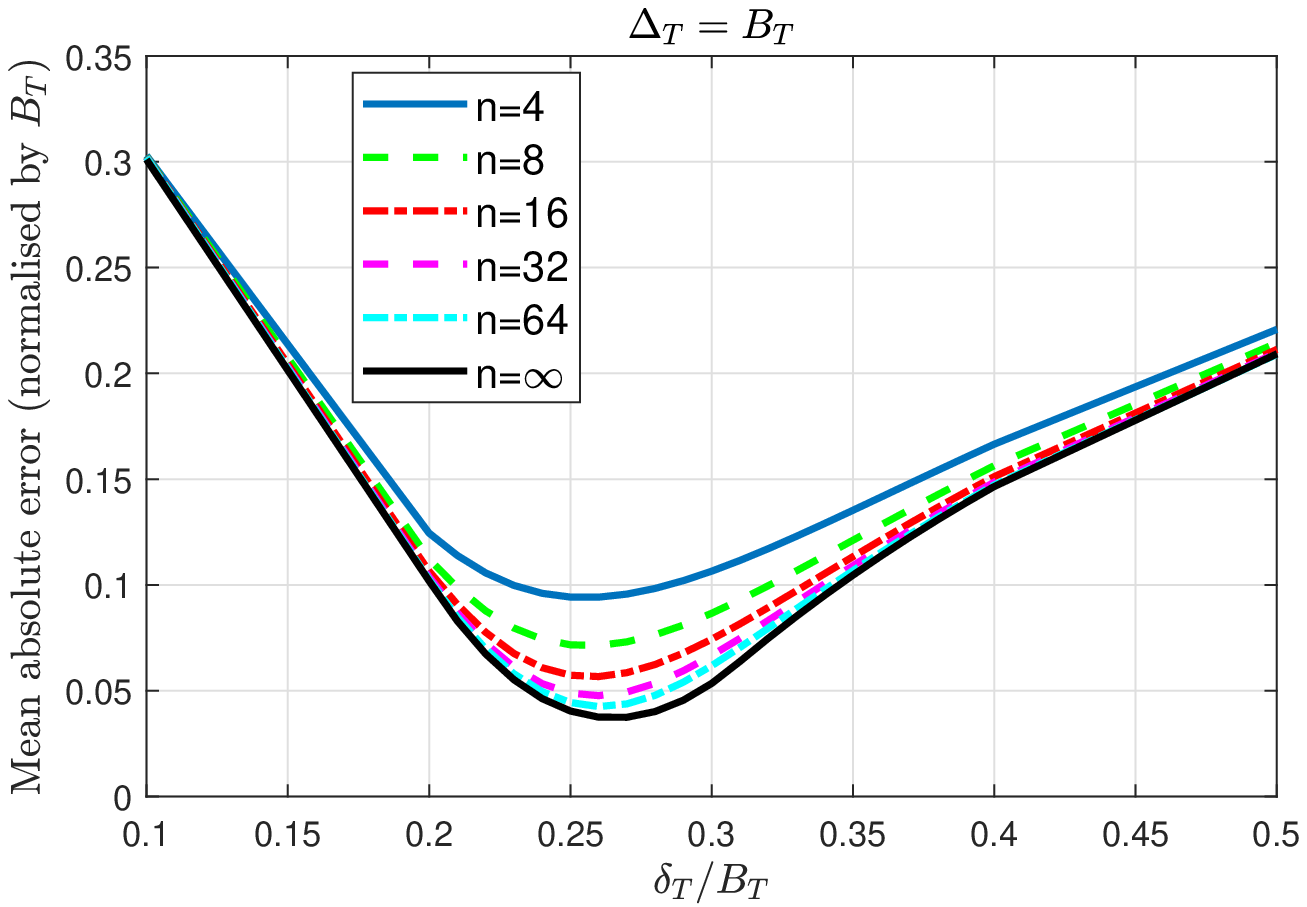}	\\
		(a)	
	\end{minipage}
	\begin{minipage}{.48\linewidth}
		\centering
		\includegraphics[width=1\textwidth]{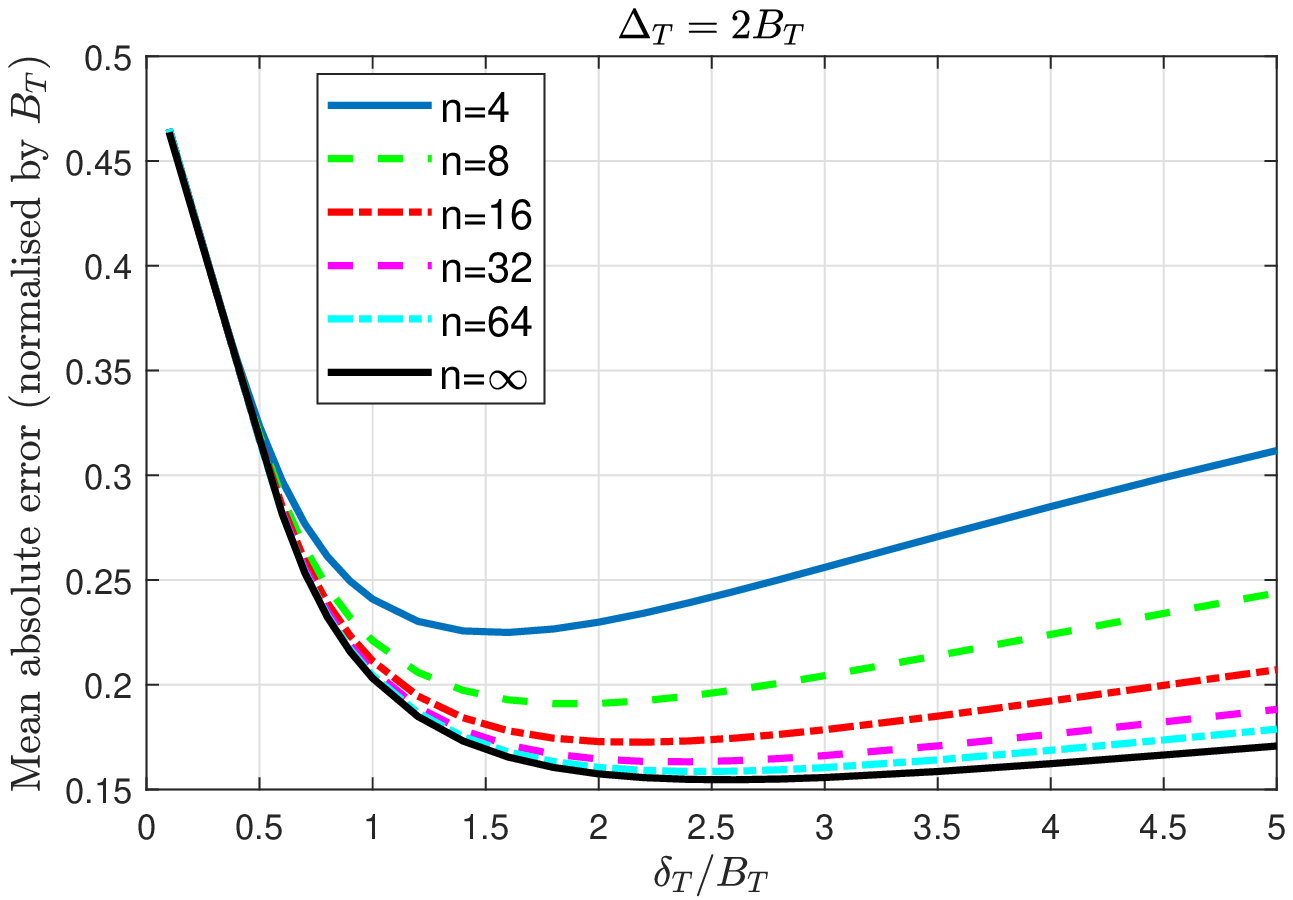}\\
		(b)
	\end{minipage}
	\caption{Mean absolute error $\mathbb E\{|\epsilon'_{\psi}(t)|\}$: $\gamma=-10$ dB, $N_T=64$, $N_R=1$.}\label{Fig:mean_error_delta_Delta}
	\vspace{-1.5em}
\end{figure}

Fig.~\ref{Fig:mean_error_delta_Delta} plots the MAE normalised by $B_T$, i.e., $\mathbb E\{|\epsilon'_{\psi}(t)|\}/B_T$, as a function of $\delta_T/B_T$, for a number of values of the pilot sequence length $n$, when $\Delta_T=B_T$ (Fig.~\ref{Fig:mean_error_delta_Delta} (a)) and $\Delta_T = 2B_T$ (Fig.~\ref{Fig:mean_error_delta_Delta} (b)). The results are obtained for the one sided case, where only the BS has multiple antennas and UE has an omni antenna. In this case, $\epsilon_R=0$.

As can be seen from Fig.~\ref{Fig:mean_error_delta_Delta}, the best $\delta_T$ that minimises the MAE, i.e., $\delta_T^* = \arg\min_{\delta_T}E\{|\epsilon'_{\psi}(t)|\}$, generally depends on the pilot sequence length $n$ and also the perturbing distance $\Delta_T$. For instance, for the sampling distance of $\Delta_T=B_T$ as shown in Fig.~\ref{Fig:mean_error_delta_Delta}(a), $\delta_T^*$ appears to be around $0.25B_T$ to $0.27B_T$ for the various values of $n$. With $\Delta_T=2B_T$, $\delta^*_T$ varies more significantly with respect to $n$, where for $n=4$, $\delta^*_T$ is around $1.5B_T$ while for $n=64$, $\delta^*_T$ is around $2.5B_T$.

\begin{figure}
	\centering
	\vspace{-2em}
	\includegraphics[width=0.55\textwidth]{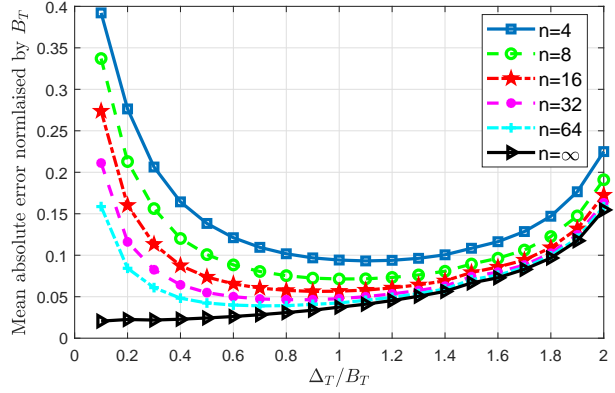}
	\vspace{-1.2em}
	\caption{Normalised mean absolute error $\mathbb E\{|\epsilon'_{\psi}(t)|\}/B_T$ versus $\Delta_T/B_T$. For each $(\Delta_T,n)$, the $\delta_T$ value that yields the smallest mean absolute error is used. \label{fig:mean_error_Delta}}
	\vspace{-1.5em}
\end{figure}

Fig.~\ref{fig:mean_error_Delta} presents the optimised MAE obtained with $\delta_T = \delta_T^*$ for a range of values on $n$ and $\Delta_T$. The curve with $n=\infty$ shows that the smaller the perturbing distance, the smaller the MAE, which appears to suggest that a smaller $\Delta_T$ is better than a larger one. However, with finite $n$ up to $n=64$, as shown by the curves in Fig.~\ref{fig:mean_error_Delta}, the MAE first decreases and then increases when $\Delta_T$ increases. The high MAE for small $\Delta_T$ (e.g., $\Delta_T=0.1B_T$) when $n$ is small results from the difficulty in differentiating two beams that are pointing in similar directions and thus have comparable and high gains. The high MAE for large $\Delta_T$, e.g., $\Delta_T=2B_T$, comes with a similar difficulty, which is caused by two beams that are both far apart from the dominant path and hence have low gains.

Fig.~\ref{fig:mean_error_Delta} also shows that the best perturbing distance varies with $n$: for a larger $n$, the best perturbing distance is smaller and vice versa. For instance, for $n=8$, the best perturbing distance $\Delta_T$ is around $B_T$. When $n=64$, the best perturbing distance becomes close to $0.6B_T$. However, for $n$ up to $64$, it appears that $\Delta_T = B_T$ is a robust choice as the corresponding MAE is close to the minimum MAE that can be achieved by other values of $\Delta_T$. Recall that in Fig.~\ref{Fig:mean_error_delta_Delta}(a), the best step size is $\delta_T^*\approx B_T/4$ when $\Delta_T=B_T$. Therefore, $(\Delta_T = B_T,\delta_T = B_T/4)$ is a good parameter choice.

The fact that $(\Delta_T = B_T,\delta_T = B_T/4)$ is a good parameter choice does not rule out other choices also being good choices. For instance, $(\Delta_T = B_T/2, \delta_T = B_T/3)$ has similar MAEs to $(\Delta_T = B_T,\delta_T = B_T/4)$. With this observation, we also consider the PLT that captures the performance in dynamic scenarios, in order to further optimise the parameter choices.
	
\subsubsection{Probability of Losing Track between two tracking corrections}\label{Sec:PLT}
In this subsection, we will discuss the PLT between any two consecutive tracking corrections where the initial error before the first correction is within $[-B_T,B_T]$. Let $t$ and $t+T$ be the slot indices of the two consecutive tracking slots. We also consider that the angular change between the start of slot $t$ and the start of slot $t+T$ is Lipschitz continuous in time.

As before, we denote the initial beam alignment error before the correction, by $\epsilon_{\psi(t)}$. Let $\tilde{h}$ denote the random correction as obtained from~\eqref{eqn_control}, $\delta_{\psi}(\tau) = \psi(t+\tau)-\psi(t)$ the angular change and $\tau \in [0,T)$. A loss of track occurs either when the error immediately after correction goes outside the interval $[-B_T,B_T]$, i.e., ${\cal A}_{\epsilon_{\psi(t)}} = \{\tilde{h}: \abs{ \epsilon_{\psi}(t) + \tilde{h} }> B_T\}$, or the error goes outside the interval at an intermediate time $\tau\in[t,t+T]$, i.e., $\abs{ \epsilon_{\psi}(t) + \tilde{h}+\delta_{\psi}(\tau) }> B_T$, but ${\cal A}_{\epsilon_{\psi(t)}}$ does not occur. Since the angular change is Lipschitz continuous, there exists a constant $a\geq 0$ such that $\abs{\delta_{\psi}(\tau)}\leq a$. Therefore, a necessary condition to $\abs{ \epsilon_{\psi}(t) + \tilde{h}+\delta_{\psi}(\tau) }> B_T$ is that $\abs{ \epsilon_{\psi}(t) + \tilde{h}} + a> B_T$, since $\abs{ \epsilon_{\psi}(t) + \tilde{h}} + a\geq\abs{ \epsilon_{\psi}(t) + \tilde{h}}+\left|\delta_{\psi}(\tau)\right|\geq\abs{ \epsilon_{\psi}(t) + \tilde{h}+\delta_{\psi}(\tau) }$. Denote
${\cal B}_{\epsilon_{\psi(t)}}(a) = \{\tilde{h}:\abs{ \epsilon_{\psi}(t) + \tilde{h}} + a> B_T\}$. Thus, the probability of
exiting the interval $[-B_T, B_T]$, either initially or during the update interval, over all possible initial errors, can then be upper bounded by
\begin{align}
J_a =  \sup_{\epsilon_{\psi}(t) \in [-B_T, B_T] } \left[\prob{{\cal A}_{\epsilon_{\psi(t)}}}+\prob{{\cal B}_{\epsilon_{\psi(t)}}(a)\bigcap\overline{{\cal A}_{\epsilon_{\psi(t)}}}}\right],
\end{align}
where $\overline{{\cal A}_{\epsilon_{\psi(t)}}}$ denotes the complementary event to ${{\cal A}_{\epsilon_{\psi(t)}}}$.

The PLT bound $J_a$ depends on $a$, which reflects the maximum angular change during the update interval. (For fixed update interval, the parameter $a$ reflects the speed of angular change.) $J_a$ also depends on other algorithm parameters including $\Delta_T$, $\delta$ and the pilot sequence length $n$. Moreover, the bound is tight in the sense that given $J_a$, we can find an initial error $\epsilon_{\psi(t)}\in [-B_T, B_T]$ such that the PLT is arbitrarily close to $J_a$. In other words, the bound represents the worst case PLT starting from an arbitrary initial error.

To further understand the impact of mobility on PLT, we present the following lemma.
\vspace{-0.5em}
\begin{lemma}\label{lemma_1}
	Under the fixed single-path model, $J_a$ is monotonic increasing for $a \in [0,B_T)$.
\end{lemma}

The proof of Lemma~\ref{lemma_1} follows immediately from the definitions of $J_a$ and ${\cal B}_{\epsilon_{\psi(t)}}(a)$ and because the $\sup$ is being taken.
This lemma shows that a larger $a$ caused by a higher mobility and/or a longer time duration between two updates (thus a lower tracking rate), will lead to a higher $J_a$. Our numerical evaluations of $J_a$ show that for moderate or high values of $a$, e.g., $a\geq 0.4B_T$, $J_a$ is dominated by $\left\{{\cal B}_{\epsilon_{\psi(t)}}(a)\bigcap\overline{{\cal A}_{\epsilon_{\psi(t)}}}\right\}$ and that $J_a$ is sensitive to the choices of $(\Delta_T, \delta_T)$ as well as the pilot sequence length $n$. Thus it should be used to further determine these parameter choices.

	
 \begin{figure}
 	\vspace{-2em}
\begin{minipage}{.48\linewidth}
	\centering
	\includegraphics[width = 0.95\textwidth]{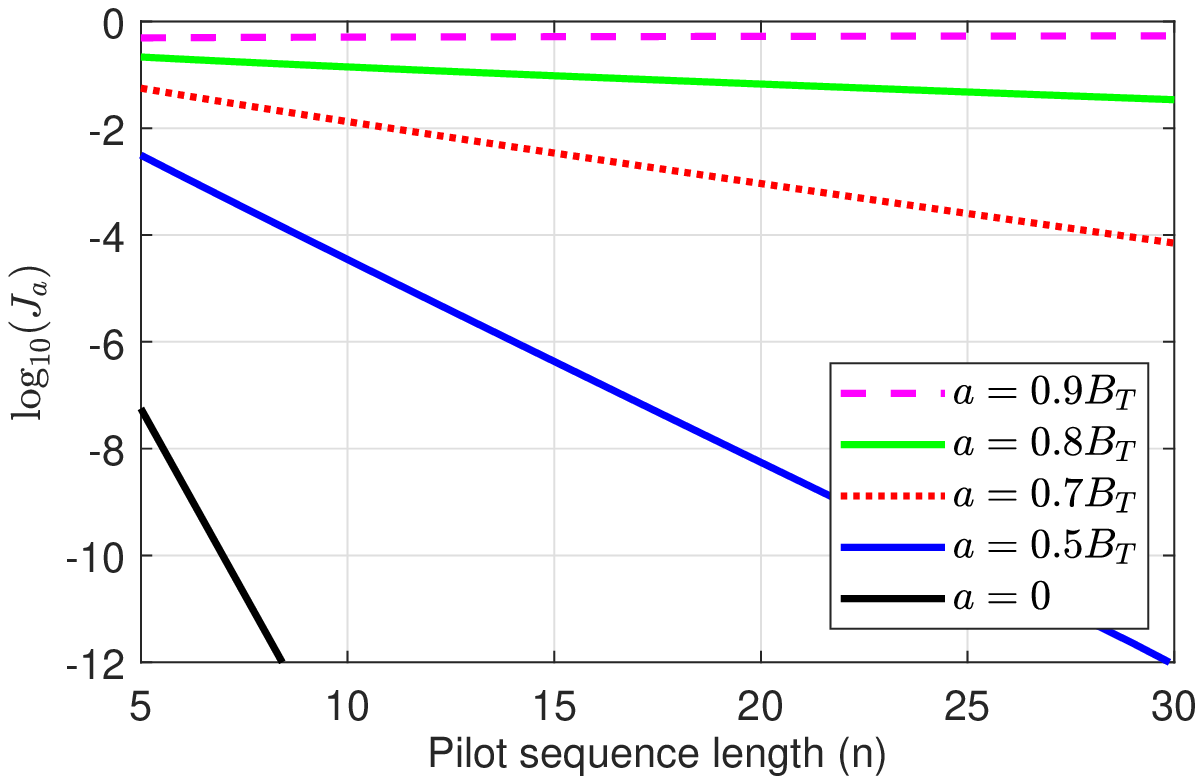}
	\vspace{-1em}
	\caption{Probability bound of exiting $[-B_T, B_T], \Delta_T = B_T, \delta_T = B_T/4$ with $\gamma = -10 dB$. \label{fig_JumpSamp}}
\end{minipage}
\hspace{1em}
\begin{minipage}{.48\linewidth}
	\centering
	\includegraphics[width = 0.95\textwidth]{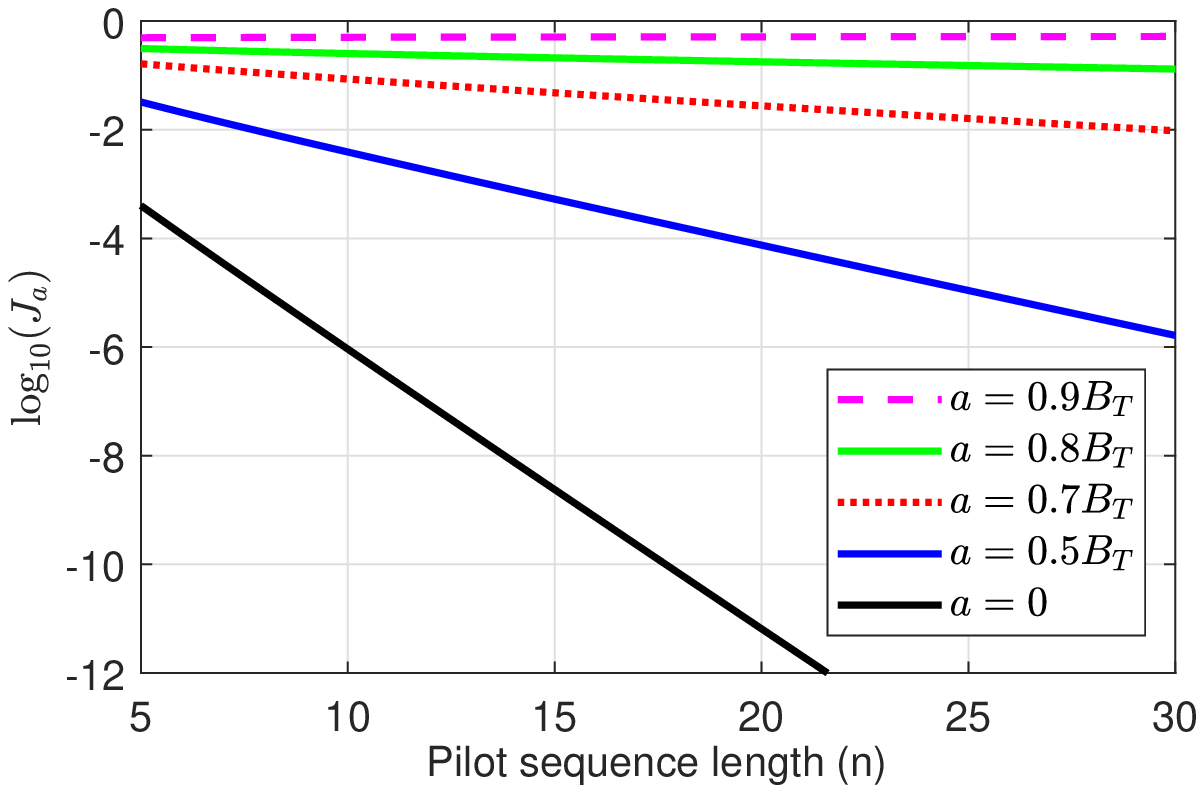}
	\vspace{-1em}
	\caption{Probability bound of exiting $[-B_T, B_T], \Delta_T = B_T/2, \delta_T = B_T/3$ with $\gamma = -10 dB$. \label{fig_JumpBthirdSamp}}
\end{minipage}
\vspace{-1.5em}
\end{figure}

Fig.~\ref{fig_JumpSamp} and Fig.~\ref{fig_JumpBthirdSamp} graph $J_a$ for two different pairs of $(\Delta_T,\delta_T)$, i.e., $(\Delta_T = B_T,\delta_T = B_T/4)$ and $(\Delta_T = B_T/2,\delta_T = B_T/3)$, respectively. From both figures, it can be seen that a higher $a$ leads to a higher $J_a$ for the same $n$. Also $J_a$ decays as the pilot sequence length $n$ increases. Moreover, by comparing Fig.~\ref{fig_JumpSamp} with Fig.~\ref{fig_JumpBthirdSamp}, it can be seen that for the same $(n,a)$, $J_a$ for $(\Delta_T = B_T,\delta_T = B_T/4)$ is lower than that for $(\Delta_T = B_T/2,\delta_T = B_T/3)$. This means that for the same tracking overhead, parameter choice $(\Delta_T = B_T,\delta_T = B_T/4)$ leads to lower PLT than $(\Delta_T = B_T/2,\delta_T = B_T/3)$, thus is a better choice.

Overall, through the MAE and PLT analysis above, it can be concluded that the choice of $(\Delta_T =  B_T,\delta_T =  B_T/4)$ for the SC algorithm provide lower mean absolute error after tracking update and lower PLT. Therefore, in the simulations in Section~\ref{Sec:Results}, we use this particular parameter combination. We note that the suggested parameter combination has taken into account the number of antennas through $B_T = 1/N_T$ and hence is applicable to ULAs with any number of antennas. For each tracking update, it is required to compute \eqref{eq:track_update}, whose computational complexity is dominated by computing $Q_R$ and $Q_T$ given in \eqref{eq_Qr_Qt} and is on the order of $2n$.

\begin{remark}
	 The results presented in Fig.~\ref{fig_JumpSamp} show that $J_a$ can be reduced by 1) increasing $n$ with $a$ fixed and 2) decreasing $a$ with $n$ fixed. Since $a$ can be reduced by performing tracking updates more frequently, i.e., increasing the tracking overhead, and increasing $n$ also increases the tracking overhead, it can be concluded that $J_a$ generally decreases as the overhead increases. Increasing the overhead, however, will reduce the available time for data transmission. Thus, as expected, there are tradeoffs between the tracking overhead and the system performance.
	
	 In fact, there are also tradeoffs between the tracking rate, or equivalently $a$, and the pilot length $n$, since the same amount of overhead can be achieved by different combinations of $a$ and $n$. Our analysis in Appendix~\ref{Append:tradeoff}, assuming an idealised scenario where the angular change speed is constant and known and there is no fading, shows that it is important to control the tracking rate such that $a$ is neither too small nor too large. The results in Appendix~\ref{Append:tradeoff} show that moderate $a$, e.g., in the range of $(0.3B_T,0.6B_T)$, requires a minimum amount of overhead to meet a performance target measured by the probability of losing track over $T$ slots. Setting the tracking rate very high or very low such that $a$ is very small or very large will increase the amount of overhead required. This is because for smaller $a$, the required pilot length tends to decrease more slowly and hence the increased tracking rate (as a result to keep $a$ smaller) becomes the dominant factor of the overhead. This effect becomes obvious when $a=0.1B_T$. Very large $a$ such as $0.8B_T$ requires very long pilots which pushes up the overhead, as shown in Table~\ref{table:tradeoff} in Appendix~\ref{Append:tradeoff}. Furthermore, $a=0.7B_T$ is about the maximum angular change that can be tracked reliably with reasonably low overhead.
	
	 Nevertheless, the tradeoff analysis in Appendix~\ref{Append:tradeoff} shows the importance of adjusting the tracking rate such that $a$ is always in the right range, according to the time-varying angular change speed. For this reason, in Section~\ref{Sec:adaptive_rate}, we present a method to set the tracking rate adaptively based on estimates of the angular change rate. $\square$
\end{remark}
	
\vspace{-0.5em}	
\section{The Algorithm of Adaptive Tracking with Stochastic Control (ATSC)}\label{Sec:Algorithm}
In Section~\ref{Sec:Control}, we have presented the stochastic control algorithm for beam tracking and discussed the parameter choices. For the complete ATSC framework as explained in Section~\ref{Sec:System}, there are two additional elements: (1) realignment and (2) adaptive tracking frequency. In what follows, we elaborate these elements in order.

\vspace{-1em}		
\subsection{Realignment}
\vspace{-0.5em}
We begin by presenting the criteria of realignment. A realignment is needed when the path to be tracked is lost or blocked or disappears. In these circumstances, the SNR after applying the data beams may degrade significantly compared to the previous time slots when the path is on track. Hence, it is reasonable to  initiate a realignment when the SNR after beamforming has dropped significantly compared to a reference level. To formally present our adopted criteria of realignment, let $t_0<t$ be the latest slot that the beams were updated (by either a tracking update or a realignment) before slot $t$ and ${\cal{T}}(t) = \{t_0,t_0+1,\ldots,t-1\}$ be the set of slots prior to slot $t$. A realignment is initiated at slot $t$ if
{\small
\begin{equation}\label{eq:relignment}
\Gamma(t) < \max_{\tau \in {\cal T}(t)} \Gamma(\tau) - \zeta,
\end{equation}}
where $\zeta>0$ is a threshold in dB and $\Gamma(t)$s are also measured in dB.

Once the realignment condition is triggered, the BS and the UE will attempt to re-discover the dominant path  at slot $t+1$ via methods such as~\cite{liu2017Jsac} or~\cite{min2019TWC}. There can be possible refinements to these methods to better meet the need for realignment, which are left for future works.

\vspace{-1em}
\subsection{Adaptive Tracking Rate}\label{Sec:adaptive_rate}
\vspace{-0.5em}
It is expected that when the angle of the dominant path changes fast, tracking measurements are to be collected more frequently, and vice versa. With this purpose, the tracking rate is set adaptively according to the angular speed estimated based on beam angle updates. The algorithm works as follows.

Denote ${\cal T}_f(t) = \{t,t-t_f,t-2t_f,\ldots,t-(T_f-1)t_f\}$ as the set of slots that $T_f$ tracking updates are performed, where $t_f\geq 1$ is the number of slots per tracking interval. At each tracking slot $\tau \in {\cal T}_f(t)$, the BS and/or the UE angles are updated. The algorithm uses these updated angles to estimate the angular change rate per slot:
\begin{equation}\label{eq:psi_est}
\alpha_{\psi} =\sum^{T_f-2}_{k=0}\frac{|\Psi(t-kt_f)-\Psi(t-(k+1)t_f)|}{t_f(T_f-1)} ~~\text{and}~~
\alpha_{\phi} = \sum^{T_f-2}_{k=0}\frac{|\Phi(t-kt_f)-\Phi(t-(k+1)t_f)|}{t_f(T_f-1)}.
\end{equation}
Suppose the targeted angular change per tracking intervals are $\beta_T$ and $\beta_R$ at the BS and the UE, respectively. Then the expected number of slots per tracking interval, i.e., $t_f$, can be updated as: $t_f\leftarrow \lfloor \frac{\beta_T}{\alpha_{\psi}} \rfloor $ and $t_f\leftarrow \lfloor \frac{\beta_R}{\alpha_{\phi}} \rfloor $ for the BS and the UE, respectively. As we assume that the BS and UE tracking are performed at the same rate and at the same slots, $t_f$ can be updated as:
\begin{equation}\label{eq:track_f}
t_f \leftarrow \max \lc \min\lb \lfloor \frac{\beta_T}{\alpha_{\psi}} \rfloor, \lfloor \frac{\beta_R}{\alpha_{\phi}} \rfloor  \rb, 1 \rc.
\end{equation}
Note that $t_f$ is calculated according to the targets $\beta_T$ and $\beta_R$, which must be chosen to be consistent to the tracking capability of the SC algorithm and according to the beam width. As we have shown in Appendix~\ref{Append:tradeoff}, angular change up to $0.7B_T$ per tracking interval is about the limit that can be tracked reliably by the SC algorithm with reasonably low overhead and accurate information of the angular change rate. We therefore will consider in Section~\ref{sec:two_sided} $\beta_T = 0.7B_T$ and $\beta_R=0.7B_R$ when testing the ATSC. We will also consider lower targeted values of $\beta_T$ and $\beta_R$, which will give ATSC more tolerance to the errors of the angular speed estimates. We finally summarise ATSC in Table~\ref{table:algorithm}.

\begin{table}[t]
	\begin{center}
		\vspace{-2em}
		\caption{Algorithm Summary for ATSC} \label{table:algorithm}
		\vspace{-1em}
		\resizebox{0.5\textwidth}{!}{\begin{tabular}{l}
				\hline
				{\bf Input}: $\Delta_T$, $\Delta_R$, $n$, $\beta$, $\zeta$, $T_f$ \\
				{\bf Initialisation}: $t_f \leftarrow 1$, $c_f \leftarrow 1$, \\
				 flag\_dis $\leftarrow 0$, perform initial alignment\\
				$t \leftarrow 1$ \\
				{\bf Loop over slot $t$}\\
				\quad {\bf If flag\_dis = 1}: Perform a full search (realignment) \\
				\quad {\bf Else}: \\
				\quad\quad\quad \quad {\bf If $c_f\geq t_f$}: Perform tracking \\
				\quad\quad\quad \quad\quad\quad Collect measurements $Q_T^{\pm}(t)$ and $Q_R^{\pm}(t)$ \\
				\quad\quad\quad \quad\quad\quad Update angle $\Psi(t)$ and $\Phi(t)$ according to \eqref{eq:track_update}\\
				\quad\quad\quad \quad\quad\quad Update $t_f$ according to~\eqref{eq:track_f} \\
				\quad\quad\quad \quad\quad \quad $c_f\leftarrow 1$\\
				\quad\quad\quad \quad {\bf Else}: \\
				\quad\quad\quad \quad\quad\quad $c_f\leftarrow c_f+1$ \\
				\quad Estimate the post-beamforming SNR $\Gamma(t)$ \\
				\quad {\bf If} reailgnment criteria~\eqref{eq:relignment} is met: \\
				\quad\quad flag\_dis $\leftarrow 1$ \\
				\quad {\bf Else}: flag\_dis $\leftarrow 0$ \\
				\quad$t\leftarrow t+1$ \\
				\hline
		\end{tabular}}
	\end{center}
	\vspace{-3 em}
\end{table}

\vspace{-1 em}
\section{Experiments and Discussions}\label{Sec:Results}
In this section, we consider three experiments to evaluate the performance of the proposed ATSC framework for mmWave beam tracking. In the first experiment, we consider a single-sided beam tracking problem where only the BS has multiple antennas (hence requiring beam tracking). We use this experiment to examine the stochastic control (SC) algorithm for beam tracking as detailed in Section~\ref{Sec:Control}, further verify our analysis and make comparisons to one state-of-the-art algorithm in the literature. In the second experiment, we consider more complicated scenarios where both the BS and the UE have multiple antennas, hence both requiring beam tracking, and also consider fading channels. We use this experiment to test the proposed ATSC framework (with adaptive tracking frequency and realignment) and to determine the tracking targets $\beta_T$ and $\beta_R$. In the third experiment, we evaluate the performance of ATSC in a more realistic scenario where the mmWave channels are generated using Wireless Insite~\cite{Insite}, a commercial ray-tracing simulator that is known for its high accuracy and has been widely used in the literature~\cite{yang2006propagation,li2018generative,va2017inverse}.


\vspace{-1em}
\subsection{One-Sided Tracking with Fixed Single-Path Channel\label{sec_Expt1}}
\vspace{-0.5em}
In this experiment, BS has $N_T=64$ antennas and UE has $N_R=1$ antennas, thus only the BS angle is to be tracked. The UE is moving around the BS along a circle centred at the BS such that the SNR without beamforming at the UE is fixed at $-10$ dB and the angle is changing at a constant speed such that $|\psi(t)-\psi(t-1)|=c$, where $c\geq 0$ is the constant angular speed. The channel has only one path with fixed channel gain.

For each trial of the experiment, we simulate $T=1,000$ slots, within each of which the angle is assumed fixed. Tracking measurements and beam updates are performed every 10 slots, e.g., at slot $\{1, 11, 21, \ldots\}$. In this case, the angular change per tracking interval is $\beta_T = 10c$.

The initial error at the start of each trial, i.e, at slot 1, is assumed uniformly distributed in $[-B_T,B_T]$, where we remind the reader that $B_T = 1/64$ is half of the ULA beam width. This initial error corresponds to the error after an initial beam alignment via scanning a fixed set of beams~\cite{min2019TWC}. We repeat the trials for 10,000 times to collect summary statistics. The results of the proposed SC algorithm are compared to the tracking algorithm developed in~\cite{zhu2018high}, which also uses signals collected from two sampling beams to update the beam. Note that a perturbing distance of $\Delta_T=2B_T$ was suggested by~\cite{zhu2018high} while the SC algorithm adopts $\Delta_T=B_T$ based on the MAE and PLT analysis. The algorithm in~\cite{zhu2018high} also has a different updating mechanism.

For each slot $t$, the post-beamforming SNR $\xi(t)$ is calculated as:
$\xi(t) = \left|\mathbf{w}^\dag(t)\mathbf{H}(t)\mathbf{f}(t)\right|^2/\sigma^2$,
where $\sigma^2$ is the variance of the effective noise. Let $\mathbf{w}^*(t) = \mathbf{w}(\psi^*(t))$ and $\mathbf{f}^*(t) = \mathbf{f}(\phi^*(t))$ be the optimal beams defined by the directions $(\psi^*(t),\phi^*(t)) = \arg\max_{\psi,\phi}\left|\mathbf{w}^{\dag}(\psi)\mathbf{H}(t)\mathbf{f}(\phi)\right|^2$.~Clearly,  the maximum achievable post-beamforming SNR at slot $t$ is $\xi^*(t) = \left|\mathbf{w}^{*\dag}(t)\mathbf{H}(t)\mathbf{f}(t)^*\right|^2/\sigma^2$.

\begin{figure}
	\centering
	\vspace{-1.5em}
	\includegraphics[width=0.52\textwidth]{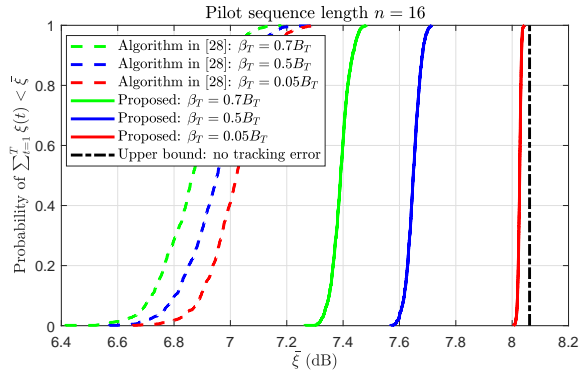}
	\caption{Empirical CDF of the average SNR after beamforming: $\prob{\frac{1}{T}\sum_{t=1}^T\xi(t)<\bar{\xi}}$. \label{fig:1sided_CDF_SNR}}
	\vspace{-1.8em}
\end{figure}

Fig.~\ref{fig:1sided_CDF_SNR} presents the empirical cumulative distribution function (CDF) of the average post-beamforming SNR, i.e., $\prob{\frac{1}{T}\sum_{t=1}^T\xi(t)<\bar{\xi}}$, for the proposed SC algorithm and the one in~\cite{zhu2018high} for different angular velocities. Pilot length is $n=16$ for both methods. Fig.~\ref{fig:1sided_CDF_SNR} also presents the empirical CDF of the post-beamforming SNR upper bound (computed assuming no tracking error) for benchmarking. It can be seen that higher angular velocities lead to larger losses of beamforming gain with respect to the maximum, for both methods. It can also be seen that the proposed algorithm provides noticeably higher average SNR than the algorithm in~\cite{zhu2018high} at all the three angular velocities. For instance, for $c = 0.05B_T$ (or equivalently $\beta_T=0.5B_T$), the proposed algorithm provides a $0.7$ dB of gain over the algorithm in~\cite{zhu2018high} ($7.65$ dB vs. $6.95$ dB).

\begin{figure}
	\centering
	\vspace{-1.8em}
	\includegraphics[width=1\textwidth]{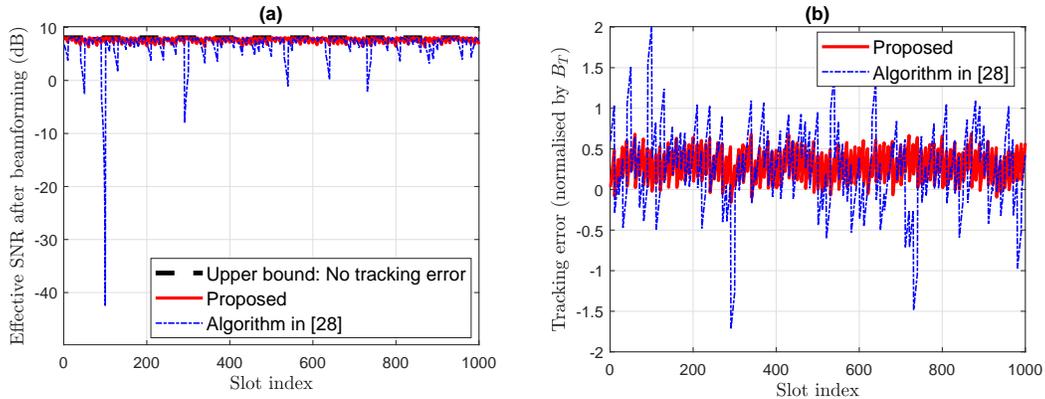}
	\vspace{-2em}
	\caption{A snapshot of the one-sided tracking experiment of SC, $\Delta_T = B_T, \delta_T = B_T/4, N_T = 64$\label{fig:1sided_snapshot} }
	\vspace{-1em}
\end{figure}

We emphasise that although the gain on the average SNR does not sound remarkable, it actually mitigates disruptive outages (due to significant losses of SNR) in the process of tracking. As a representative example, we present a snapshot of one of the 10,000 trials in Fig.~\ref{fig:1sided_snapshot}, where $c=0.05B_T$. As can be seen from Fig.~\ref{fig:1sided_snapshot}(a), the proposed algorithm maintains steady post-beamforming SNRs that are close to the upper bound. The tracking error is almost always within $[-0.5B_T,0.5B_T]$, as can be seen from Fig.~\ref{fig:1sided_snapshot}(b). The algorithm in~\cite{zhu2018high} incurred deep losses of the post-beamforming SNRs, which are disruptive and can lead to losses of data packets as well as possible need for realignment.

\begin{figure}
	\centering
	\includegraphics[width=0.52\textwidth]{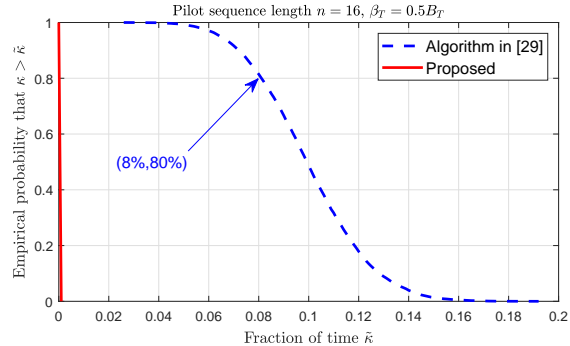}
	\vspace{-1.2em}
	\caption{One-sided experiment of SC - Complimentary CDF of the fraction of time that the tracked SNR is at least 3dB lower than the maximum SNR, i.e., $\prob{\kappa > \tilde{\kappa}}$, where $\kappa\doteq \frac{1}{T}\sum_{t=1}^T\mathbbm{1}\{[10\log_{10}(\xi(t))-10\log_{10}(\xi^*(t))]<-3\}$. \label{fig:1sided_fracT_below_optimal}}
	\vspace{-1.5em}
\end{figure}

Fig.~\ref{fig:1sided_fracT_below_optimal} shows the complimentary CDF of $\kappa\doteq \frac{1}{T}\sum_{t=1}^T\mathbbm{1}\{[10\log_{10}(\xi(t))-10\log_{10}(\xi^*(t))]<-3\}$, which is the fraction of time that the tracked SNR is at least $3$ dB lower than the maximum SNR, as measured in a single run of $T$ slots. Here $\mathbbm{1}(\cdot)$ is the indicator function and $T=1000$. In Fig.~\ref{fig:1sided_fracT_below_optimal}, $\beta_T=0.5B_T$  and $n=16$.
 It can be seen that for the algorithm from~\cite{zhu2018high}, $\kappa$ is significant for a good fraction of the simulation runs. For instance, about $80\%$ of the 10,000 runs have $\kappa > 8\%$. This means that it is common to see occurrences for which the tracked SNR is lower than the maximum possible SNR by at least 3 dB. In contrast, for the proposed algorithm, $\kappa$ is almost always zero, suggesting that such occurrences are very rare.

\vspace{-1em}
\subsection{Two-Sided Search with Fading Channels}\label{sec:two_sided}
\vspace{-0.5em}
In this experiment, we examine the proposed ATSC framework in a two-sided beam tracking problem, where $N_T=N_R=32$. Since $N_T=N_R$, it is natural to consider that $B_T=B_R$. For notational convenience, we denote $B_T=B_R\doteq B$ and $\beta_T=\beta_R\doteq \beta$ (the targeted angular change per tracking interval) from this subsection. The purpose of this experiment is to examine the impact of parameter $\beta$  on the overall performance of beam tracking.

In the simulations, the channel is assumed to have one dominant path and follow a Rician fading model~\cite{goldsmith2005wireless}. The BS and UE angle of the dominant path change every slot with the same velocity $c=0.05B$: $|\psi(t)-\psi(t-1)|=c$ and $|\phi(t)- \phi(t-1)|=c$. The average pre-beamforming SNR of the channel is set to $-20$ dB and the channel is generated every slot following the time-varying angle and the Rician model. Two Rician $K$-factors are considered, i.e., $13.2$ dB and $6$ dB, which correspond to LOS scenarios~\cite{Akdeniz2014} and NLOS scenarios~\cite{Akdeniz2014}, respectively. The realignment threshold is set to $\zeta=6$ dB, which is approximately the loss of beamforming gain if both the BS and UE have tracking errors around $B$.

\begin{figure}
	\centering
	\vspace{-2.2em}
	\includegraphics[width=0.95\textwidth]{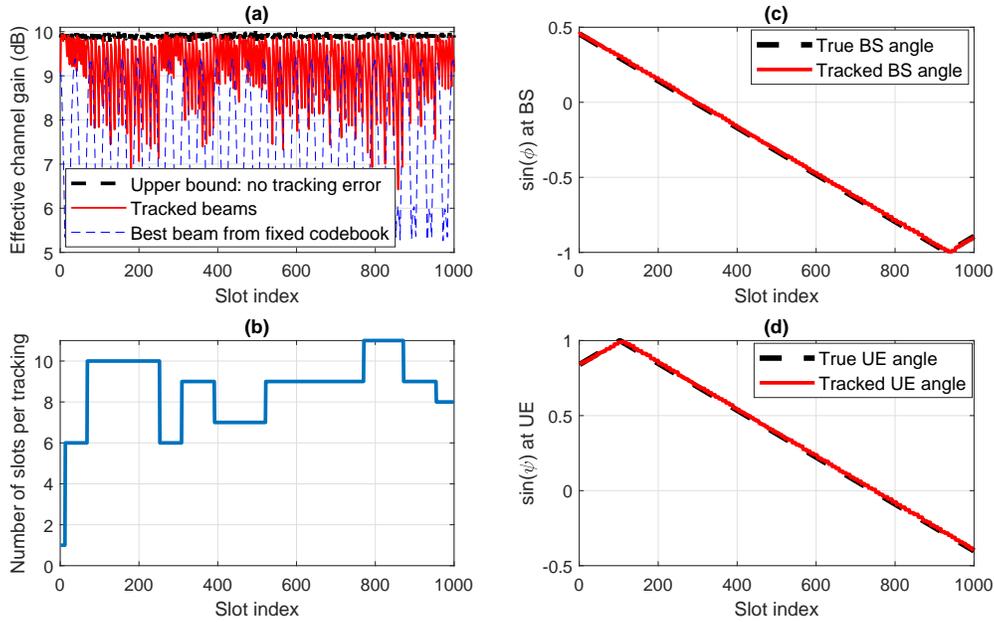}
	\vspace{-1.2em}
	\caption{Two-sided experiment of ATSC - tracking iterations snapshot: there is no realignment; $K$-factor=13.2 dB; $\beta=0.5B$.\label{fig:2sided_snapshot_1}}
	\vspace{-1.5em}
\end{figure}

Fig.~\ref{fig:2sided_snapshot_1} presents the results of one typical simulation run (of 1,000 slots) in LOS with $n=16$ and $\beta=0.5B$. Fig.~\ref{fig:2sided_snapshot_1} (a) presents the SNR of the tracked channel (the red curve), the SNR upper bound achieved by the optimal directional beamforming (the black curve) and a baseline SNR (the blue curve) achieved by the best BS-UE beam pair from fixed codebooks that have $N_T$ and $N_R$ DFT beams. Fig.~\ref{fig:2sided_snapshot_1} (b) presents the number of slots per tracking interval, i.e., $t_f$ in~\eqref{eq:track_f}. Fig.~\ref{fig:2sided_snapshot_1} (c) and (d) presents the directions of the true path and the tracked beam at Tx and Rx, respectively.

From Fig.~\ref{fig:2sided_snapshot_1} (a), it can be seen that the tracked beam produces SNR that is almost always within 3 dB of the upper bound and thus does not meet the realignment threshold $\zeta=6$ dB. Thus there is no realignment. This performance can be explained by the results in Fig.~\ref{fig:2sided_snapshot_1} (c) and (d), which show that the tracked BS/UE angles are close to the true BS/UE angles throughout the run. It can also be seen that the SNR of the tracked channel is higher than the baseline for most of the time, which further demonstrates that the tracking error is within $[-B,B]$.

Fig.~\ref{fig:2sided_snapshot_1} (b) shows that $t_f$ varies over time. The variations are due to the noisy measurements and the residual errors after tracking updates, which affect the angular velocity updates (given in~\eqref{eq:track_f}). As we mentioned earlier, the angular speed is $c=0.05B$ and $\beta = 0.5B$, hence the expected value of $t_f$ is $\beta/c = 10$. The real $t_f$ varies around this expected value, with the undesirable possibility that it can be much higher than the expected value. As we will show in the next example,  high $t_f$ values (hence less frequent tracking) due to angular velocity errors may lead to significant drops in SNR and even realignment.

\begin{figure}
	\centering
	\vspace{-2em}
	\includegraphics[width=0.95\textwidth]{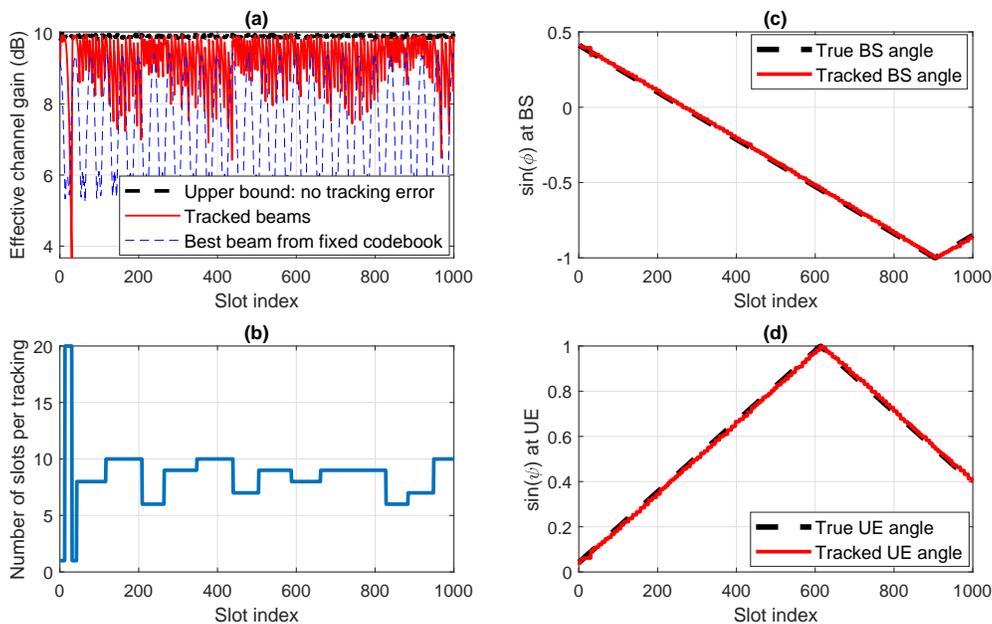}
	\vspace{-1.5em}
	\caption{Two-sided experiment of ATSC - tracking iterations snapshot: there is one realignment,$K$-factor=13.2 dB; $\beta=0.5B$.\label{fig:2sided_snapshot_2}}
	\vspace{-1.5em}
\end{figure}

Fig.~\ref{fig:2sided_snapshot_2} presents a simulation run with the same setup to Fig.~\ref{fig:2sided_snapshot_1}, but it has a realignment around slot 30. From Fig.~\ref{fig:2sided_snapshot_2}(a),  it can be seen that the tracked SNR deteriorates quickly just before slot 30. The lowest SNR (at slot 30) is below 4 dB which triggers the realignment. After the realignment, as expected, the tracking rate boosts to the maximum, i.e., tracking update is performed every single slot until sufficient number of samples to estimate the angular velocity are collected and the new tracking rate is updated. From Fig.~\ref{fig:2sided_snapshot_2}, it can be seen that the tracking rate is $t_f=20$ before slot 30, whereas the expected $t_f$ is 10 as $\beta/c = 10$. Since the angular velocity is $c=0.05B$, $t_f=20$ means that the angular change per tracking interval is $c\times t_f = B$, which already exceeds the limit of the largest angular changes that can be reliably tracked. In practice, to reduce the chance that an undesirably high $t_f$ value is adopted by the tracking algorithm (due to errors of the angular velocity estimates obtained using~\eqref{eq:psi_est}), the tracking target $\beta$ needs to be chosen more conservatively.

To see this, we present Table~\ref{table:2_sided_realignment}, which shows the fraction of simulation runs that have at least one realignment in both LOS and NLOS scenarios and for $\beta=0.5B$ and $\beta=0.7B$. As a comparison, we include results that are obtained by (i) setting $t_f$ according to estimated angular velocity and (ii) setting $t_f$ based on the true angular velocity. It can be seen that with the true angular velocity, the fraction of runs that have one or more realignment is negligible for both $\beta=0.5B$ and $\beta=0.7B$.
However, with $\beta = 0.7B$ that is close to the limit of the tracking capability, using the estimated angular velocity increases significantly the chance of realignment (caused by deep losses of SNRs). Therefore, a more conservative setup with $\beta=0.5B$ is preferable due to the much lower chance of realignment.

\begin{table}[t]
	\caption{Fraction of runs with at least one realignment. The entries in the table are the percentages of trials for which at least one re-alignment takes place. \label{table:2_sided_realignment}}
	\centering
	\scriptsize{
		\begin{tabular}{|c|c|c|c|c|}
			\hline
			& \multicolumn{2}{c}{LOS: $K$-factor = 13.2 dB} & \multicolumn{2}{|c|}{NLOS: $K$-factor = 6 dB}  \\
			\hline
			& Under estimated $c$ & Under true $c$ & Under estimated $c$ & Under true $c$ \\
			\hline
			$\beta=$0.5B & $0.32\%$ & $<0.1\%$ & $<0.21\%$   & $<0.1\%$\\
			\hline
			$\beta=$0.7B & $24.17\%$ & $<0.1\%$ & $41.56\%$ & $<0.1\%$\\
			\hline
		\end{tabular}
	\vspace{-1.5em}
	}
\end{table}

\begin{figure}
	\begin{minipage}{.49\linewidth}
		\centering
		\vspace{-2em}
		\includegraphics[width=0.95\textwidth]{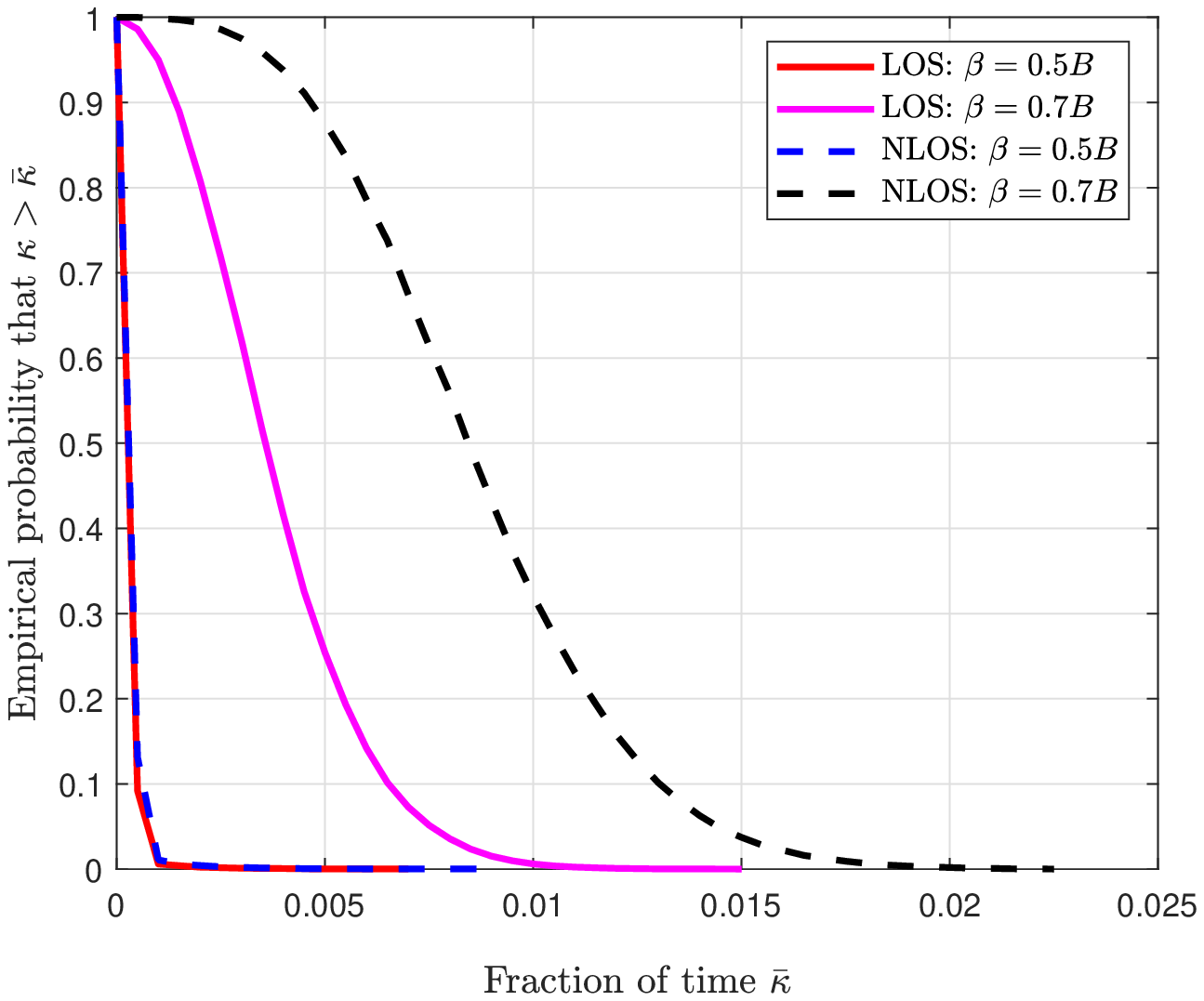}
		\vspace{-1em}
		\caption{Two-sided experiment - Complimentary CDF of the fraction of time that the tracked SNR is at least 3dB lower than the maximum SNR.\label{fig:2sided_CCDF_frac}}
	\vspace{-1.5em}	
	\end{minipage}
	\hspace{1em}
	\begin{minipage}{.49\linewidth}
		\centering
		\vspace{-2em}
		\includegraphics[width=0.95\textwidth]{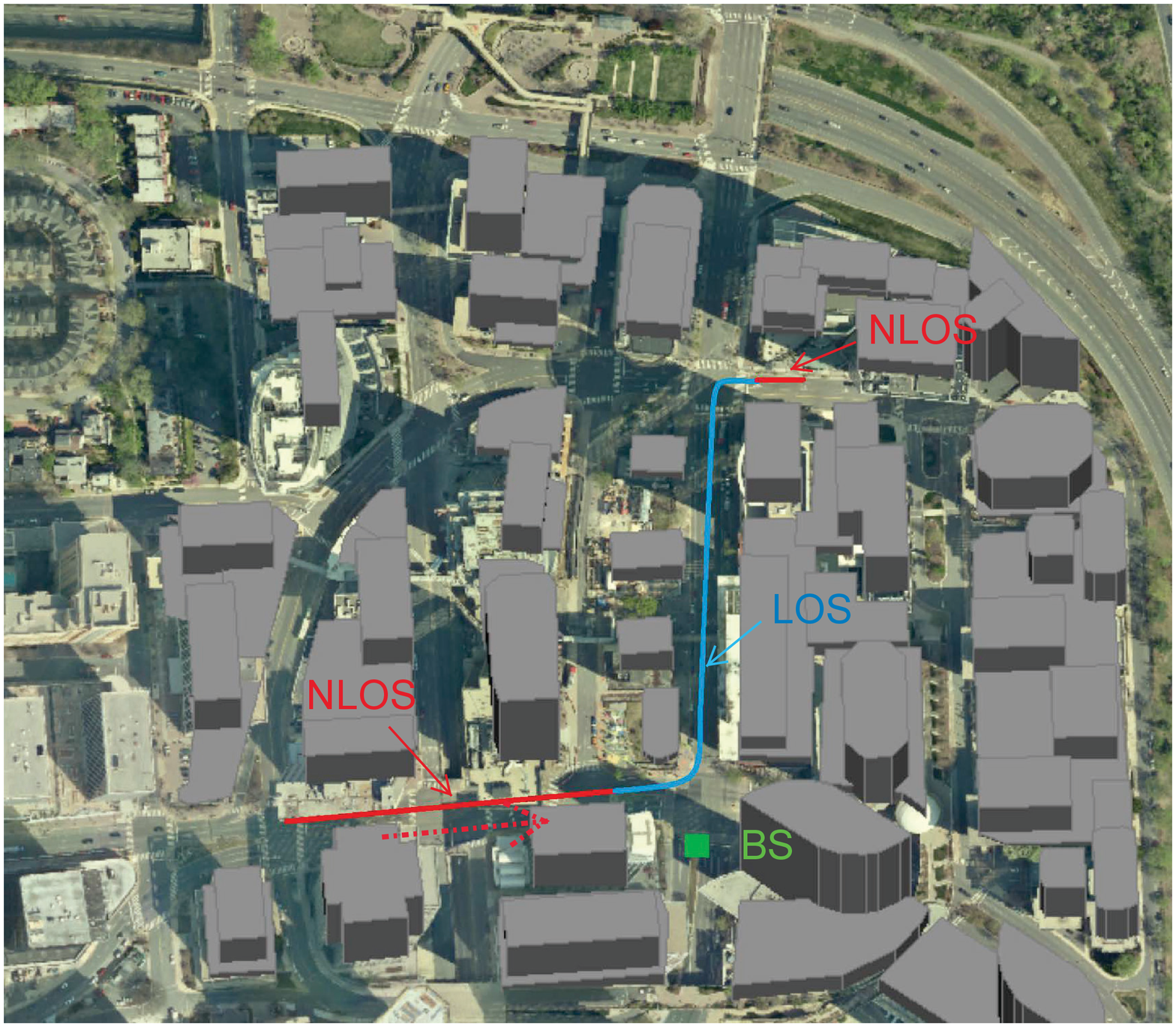}
		\vspace{-1em}
		\caption{Ray-tracing experiment: an illustration of the environment, mobile route (curve marked in red - NLOS and blue - LOS) and BS location (green square).\label{fig:ray_tracing_mobile_route} }
		\vspace{-1.5em}
	\end{minipage}
	
\end{figure}

Additionally,  we present in Fig.~\ref{fig:2sided_CCDF_frac} the complementary CDF of the fraction of time that the tracked channel is $\geq 3$ dB lower than the upper bound, i.e., $\prob{\kappa>\bar{\kappa}}$, to illustrate the impact of the choice of $\beta$. It can be seen from Fig.~\ref{fig:2sided_CCDF_frac} that when $\beta=0.5B$, the fraction of time that the tracked SNR is $\geq 3$ dB lower than the maximum SNR is close to zero for nearly all the simulation trials. However, when $\beta=0.7B$, this time fraction becomes non-negligible for many runs. For instance, for over $30\%$ of the runs in the NLOS case with $\beta=0.7B$, the time fraction is 0.01 or higher, which has triggered realignments.

%

\vspace{-1em}
\subsection{Ray-Tracing}
\vspace{-0.5em}
In this section, we examine the proposed tracking algorithm in a more realistic scenario modelled by commercial ray-tracing software~\cite{Insite}. The area considered is a 500 m-by-500 m square area in Rosslyn, Virginia, USA, as illustrated in Fig.\ref{fig:ray_tracing_mobile_route}. The considered area is a typical urban scenario with many buildings. A mmWave BS, operating at 28 GHz, is deployed in the south end of this area, as indicated by the green square in the bottom of this figure. The height of the BS is 5 m above the street level. A mmWave mobile UE is moving along the streets as illustrated by the solid line marked in red and blue in Fig.\ref{fig:ray_tracing_mobile_route}, where the two parts marked in red are NLOS and the part marked in blue is LOS. The UE has a height of 1.5 m, and the route trajectory has a total length of about 410 m, where the minimum distance between the UE and the BS is about 20 m and the maximum distance is about 240 m.

As illustrated by the red arrow in Fig.~\ref{fig:ray_tracing_mobile_route}, the UE is moving from the bottom left part of the map towards the right and then is going up. The UE is initially in a NLOS scenario where the direct path between the BS and the UE is blocked by buildings. When the UE is getting close to the street corner where the BS is deployed, the LOS path becomes available. The UE remains in the LOS scenario when it goes upwards on the map, as illustrated by the blue part of the route trajectory, until it turns to another street where the LOS path is blocked again by buildings. We note that as the distance between the UE and the BS at the very end of NLOS path is already larger than 200 m and the LOS path is not available, the UE is in fact out of the coverage range and high data rate communications cannot be supported.

In the experiment, the UE is assumed to be moving at a constant speed along the trajectory, for which we have considered pedestrian speed (4.7 km/hour), cyclist speed (20 km/hour), two typical speeds of urban vehicles (43.2 km/hour and 72 km/hour). The slot length is set to $0.5$~ms. The numbers of BS and UE antennas are set to $N_t=N_R=32$. Other simulation parameters are provided in Table~\ref{Table:Ray_tracing}. In the experiments, we use the ray-tracing software to produce path information including the BS angle, the UE angle and pathloss for a set of finely sampled points on the route trajectory, and then use linear interpolation to get the path information for all the simulation slots. After obtaining the path information, we then construct the channel matrix following~\eqref{eq:channel_model}. We note that Rician fading is added to the paths generated ($K$-factor is 13.2 dB for LOS paths and 6 dB for NLOS paths), where the path gain obtained from ray-tracing is treated as the average of the corresponding Rician variable. In the experiment, $\beta = 0.5B$, $\zeta = 6$ dB, $T_f = 10$ and the number of samples is set to $n=16$.

\begin{table}[t]
	\begin{center}
		\vspace{-1.5em}
		\caption{Parameters for the Ray-tracing Experiment} \label{Table:Ray_tracing}
		\vspace{-0.5em}
		\scriptsize{
			\begin{tabular}{|c|c||c|c||c|c|}
				\hline
				BS Height  & 5 m & Transmission Power & 30 dBm & Trajectory length & About 410 m \\
				\hline
				UE Height & 1.5 m & Thermal Noise & -99 dBm & Slot Duration & 0.5 ms \\
				\hline
				Num. Antennas ($N_T=N_R$) & 32 & Noise figure & 9.1 dB~\cite{Erricson_mmWave_noisefigure} & Mobility (km/hour) & $\{4.7, 20, 43.2, 72\}$\\
				\hline
		\end{tabular}}
	\vspace{-3em}
	\end{center}
\end{table}
\begin{figure}
	\centering
	\includegraphics[width=0.8\textwidth]{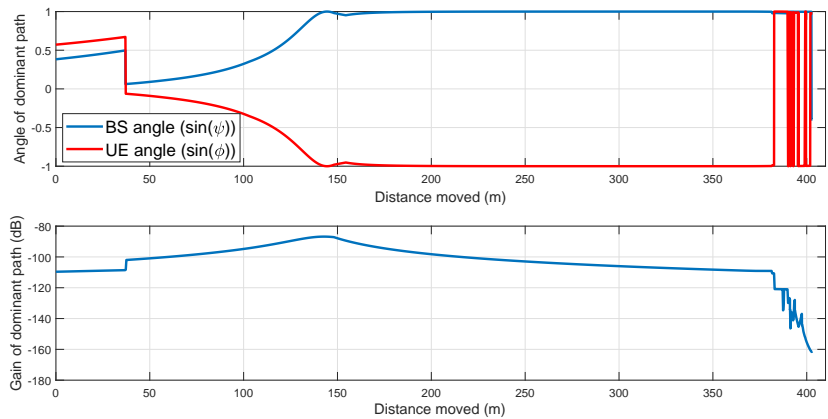}
	\vspace{-1.2em}
	\caption{Ray-tracing experiment - Dominant path: BS angle, UE angle and path gain (dB).\label{fig:ray_tracing_angle}}
	\vspace{-1.em}
\end{figure}

Fig.~\ref{fig:ray_tracing_angle} shows the BS and UE angle of the strongest paths as well as the gain of the dominant path (without beamforming) as a function of the distance moved from the beginning of the trajectory. It can be seen that both the BS and UE angle have a sudden change around 40 m. At the same position, there is a sudden increase of the gain of the dominant path. This change point corresponds to the transition from NLOS to LOS around the corner of the street where the BS is deployed, see Fig.~\ref{fig:ray_tracing_mobile_route}. From distance 380 m, the UE angle of the dominant path starts to exhibit erratic variations because it is in NLOS and the paths tend to have multiple reflections and do not last long. In the same interval, the path gain also drops significantly compared to the LOS part and is reducing very quickly.

\begin{figure}
	\centering
	\vspace{-2.5em}
	\includegraphics[width=0.9\textwidth]{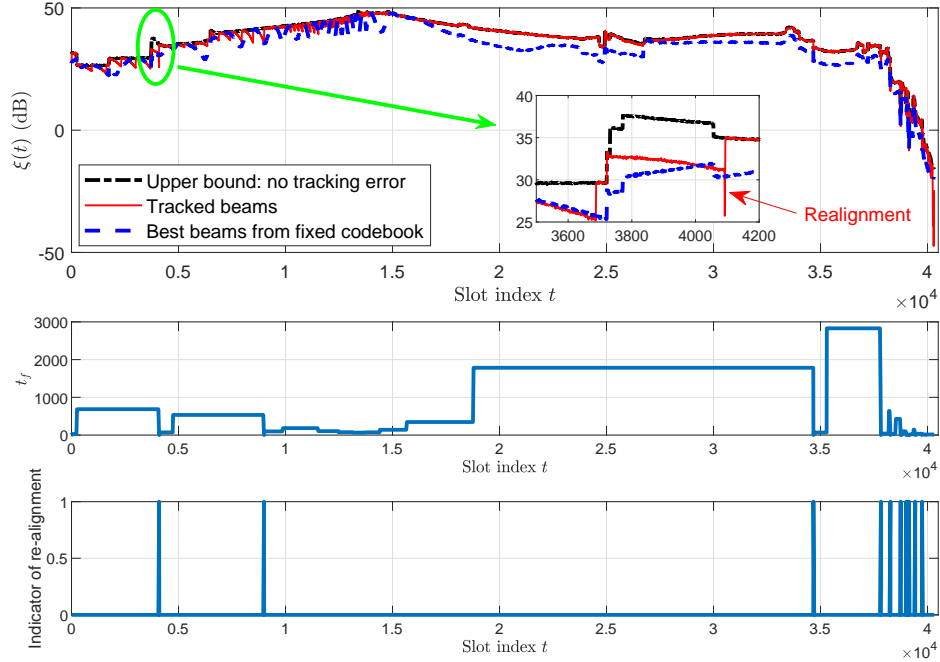}
	\vspace{-1.2em}
	\caption{A snapshot of the tracking performance for the ray-tracing experiment. UE speed: $72$ km/hour. \label{fig:snapshot_raytracing}}
	\vspace{-2em}
\end{figure}

Fig.~\ref{fig:snapshot_raytracing} graphs a snapshot plot of the tracking performance for the ray-tracing experiment, obtained with the high mobility at 72 km/hour. The first subgraph is the SNR after beamforming (the red curve), where we also include the maximum SNR obtained if no tracking error (the black curve) and the one that achieved from a fixed codebook that has $N_T=N_R=32$ DFT beams if the channel is known (the blue curve). The second subgraph presents the value of $t_f$ over the course of the simulation and the last graph presents the indicator of realignment, where '1' means there is a realignment and '0' means not.

As can be seen from the figure, until slot $0.37\times 10^4$, there is no realignment and the algorithm successfully tracked the dominant NLOS path. The tracked SNR is close to the maximum and is higher than that of the fixed codebook for most of the time. From slot $0.37\times 10^4$ where the LOS path appears, there is a short period of time (about 300 slots, 150 ms) that there is a noticeable gap between the tracked SNR and the upper bound (see the small figure inside Fig.~\ref{fig:snapshot_raytracing}(a)). This is because the NLOS path still exists and its path strength is still comparable (in fact increased considerably) to the previous period, hence the tracking algorithm keeps tracking this NLOS path, as expected. At around slot $0.41\times 10^4$, there is a realignment (see also Fig.~\ref{fig:snapshot_raytracing}(c)) because the NLOS path disappeared at that time. The realignment criteria of ATSC is triggered causing the tracking algorithm to switch to the LOS path.

Between slot $0.41\times 10^4$ and slot $3.45\times 10^4$, there is only one realignment and the tracked SNR is close to the maximum SNR nearly all the time. This demonstrates the effectiveness of the algorithm in tracking a LOS path. The one realignment at around slot $0.93\times 10^4$ is due to increases of the angular velocity at both the BS and the UE.

The realignment occurred around slot $3.5\times 10^4$ is due to the fact that the LOS path disappeared and it has to switch to a NLOS path. There are frequent realignments at the end of the trajectory (from slot $3.75\times 10^4$) as the NLOS paths become very weak and the angles vary quickly.

\begin{table}[t]
	\begin{center}
		\vspace{-1.5em}
		\caption{Overhead and tracking performance at various mobility levels in the Ray-tracking experiment.} \label{Table:Ray_tracing_summary}
		\vspace{-0.5em}
		\scriptsize{
			\begin{tabular}{|c|c|c|c|c|}
				\hline
				Mobility (km/hour) & 4.7 & 20.0 & 43.2 & 72.0 \\
				\hline
				Frac. Tracking slots & $0.34\times 10^{-3}$ & $1.65\times 10^{-3}$ & $3.25\times 10^{-3}$ & $5.2\times 10^{-3}$\\
				\hline
				$\kappa$ & $2.02\times 10^{-2}$ & $2.78\times 10^{-2}$ & $2.69 \times 10^{-2}$ & $3.24\times 10^{-2}$\\
				\hline
		\end{tabular}}
	\end{center}
	\vspace{-2.5em}
\end{table}

Fig.~\ref{fig:snapshot_raytracing} has demonstrated a successful application of the proposed ATSC framework for mmWave beam tracking. For the first 94\% of the trajectory (before  slot $3.75\times 10^4$), there are only three realignments and two of them are due to disappearances of the paths being tracked. The fraction of tracking slots is very low, at $5.2\times 10^{-3}$ (see Table~\ref{Table:Ray_tracing_summary}), despite that the UE is moving fast (72 km/hour) in this complicated urban scenario. ATSC also provides excellent tracking accuracy: $\kappa$, the fraction of time that there is a $\geq 3$ dB loss compared to the SNR upper bound, is only about 3.24\%.

For lower mobility, it has been observed that the tracking overhead, measured by the fraction of tracking slots, is even lower. For instance, as shown in Table~\ref{Table:Ray_tracing_summary}, for a cyclist speed at 20 km/hour, the overhead is only $1.65 \times 10^{-3}$ and the accuracy remains satisfactory ($\kappa=2.02\%$). Table~\ref{Table:Ray_tracing_summary} also include results for other mobility levels. These results demonstrate that the proposed ATSC can adapt to different mobility levels and achieve good tracking accuracy with low overhead.

\vspace{-0.5em}
\section{Conclusions}\label{Sec:Conclusions}
\vspace{-0.5em}
\par In this work, we have developed a new adaptive beam tracking framework called ATSC to maintain reliable links in mmWave mobile communications. ATSC is capable of tracking angular changes of the order of a large fraction of the beam width per update. It can also adaptively match the amount of tracking overhead to the time-varying angular change rate to achieve high tracking accuracy. These features have produced a tracking framework with high accuracy and reliability, as reflected by the results obtained from representative statistical channel models and realistic urban scenarios simulated by ray-tracing software. The ATSC we developed can track high mobility UE moving at 72 km/hour in complicated urban scenarios with $<1\%$ overhead.  It is expected that ATSC can be applied to track beams for multiple users in a time-division manner, owing to the small number of measurement samples required. The framework developed can also be extended to planar and other more complex arrays, although it is presented under uniform linear arrays. Possible future directions of interest include revised mechanisms for adapting tracking frequency and realignment, new tracking protocols that cope with even faster angular changes in some emerging mmWave applications with ultra-high mobility, such as high speed railways and unmanned aerial vehicles mmWave communications, and extensions to more complicated array geometries with directional antennas.

\vspace{-0.5em}
\appendices
\vspace{-0.5em}
\section{Overhead Tradeoff }\label{Append:tradeoff}
\vspace{-0.5em}
In considering overhead we will take a fixed period of $T$ slots and a probability of loss of track (PLT) $1-f$ so that $f$ is the probability that track is not lost during the $T$
slots. We consider the same setup ($N_T=64$, $N_R=1$ and pre-beamforming SNR = -10 dB) to the one used to generate Fig.~\ref{fig_JumpSamp}. Furthermore, we assume that the angular change per slot is a constant, denoted by $c$. Since $a$ is the maximum angular change per tracking interval, each tracking interval has $m_a = a/c$ slots. As $T$ is the total number of slots in a communication session, there are $N_a  = T/m_a \propto 1/a$ tracking updates in the session.

To meet the target $f$, it is sufficient that for each $a$, $[1-J_a(n_a)]^{N_a}\geq f$, where $n_a$ is the pilot length adopted and we have added the dependence of $J_a$ on $n_a$. Equivalently,
{\small\begin{equation}\label{Eq:fix_point}
J_a(n_a) \leq 1-f^{\frac{1}{N_a}}\equiv C_a ,
\end{equation}}
Let $n^*_a$ be the minimum integer satisfying Eq.~\eqref{Eq:fix_point}. Let $x$ be the number of symbols per slot, then the fraction of tracking overhead is $\frac{n^*_aN_a}{Tx} = \frac{n^*_aN_a}{N_am_ax}= \frac{n^*_a}{a/B_T}\cdot \frac{c}{B_Tx}$, which is proportional to  $\tilde{\rho}_a\doteq\frac{n^*_a}{a/B_T}$.  Clearly, for the same target $f$, $\tilde{\rho}_a$ differ according to $a$ because both the decay rate of $J_a$ with $n_a$ and $N_a$ depend on $a$. Table~\ref{table:tradeoff} shows results where  the performance target is set to $f=0.95$ and $N_a$ is obtained by rounding to the nearest integer value.

As shown in Table~\ref{table:tradeoff}, $a=0.5B_T$ requires the minimum amount of overhead.  Larger or smaller values of $a$ all require higher overhead.  For smaller $a$, this is because $N_a$ is much larger whereas the number of pilots required, i.e., $n_a^*$, is almost the same. For very large $a$, the overhead is dominated by the large $n^*_a$.

Note that $a=0.5B_T$ cannot be taken as the ultimate optimal choice of tracking target, because it is obtained based on the chosen target $f=0.95$. Nevertheless we find through similar calculations with different $f$ and $T$ that moderate $a$ in the range $\in (0.3B_T, 0.6B_T)$ offers good efficiency. The higher choice $a=0.7B_T$ is about the maximum angular change that can be tracked reliably and requires larger overhead to do so. More importantly, as the angular change speed $c$ has to be estimated in practice, pushing the tracking target towards the limit $0.7B_T$ can lead to frequent loss of track and thus re-alignment (see the results in Table II).

\begin{table}
	\center
	\vspace{-1.5em}
	\caption{Tradeoff analysis for $\Delta_T=B_T$, $\delta_T = B_T/4$: $f=0.95$.}\label{table:tradeoff}
	\vspace{-0.8em}
	\scriptsize{
	\begin{tabular}{|c|c|c|c|c|c|c|c|}
		\hline
		$a$ & $0.1B_T$ &$0.2B_T$ &$0.3B_T$ & $0.4B_T$&$0.5B_T$ &$0.6B_T$ &$0.7B_T$\\
		\hline
		$N_a$&100 & 50 & 33 & 25 & 20 & 17 & 14 \\
		\hline
		Pilot length $n^*_a$ & 3 &  3 &  4 & 5 & 6 & 8 & 15 \\
		\hline
		$\tilde{\rho}_a$ & 30 & 15 & 13.3 & 12.5 & 12  & 13.3 & 21.4 \\
		\hline
	\end{tabular}
}
\vspace{-2.5em}
\end{table}

\vspace{-0.5em}
\bibliographystyle{IEEETran}
\bibliography{IEEEabrv,Reference_mmWaveRev}

\end{document}